\documentclass[fleqn,usenatbib]{mnras}
\pdfoutput=1
\usepackage[T1]{fontenc}
\usepackage{ae,aecompl}

\usepackage{geometry}
\usepackage{graphicx}
\usepackage{amsmath}
\usepackage{amssymb}
\usepackage{array}

\makeatletter
\newcommand*\rel@kern[1]{\kern#1\dimexpr\macc@kerna}
\newcommand*\widebar[1]{%
  \begingroup
  \def\mathaccent##1##2{%
    \rel@kern{0.8}%
    \overline{\rel@kern{-0.8}\macc@nucleus\rel@kern{0.2}}%
    \rel@kern{-0.2}%
  }%
  \macc@depth\@ne
  \let\math@bgroup\@empty \let\math@egroup\macc@set@skewchar
  \mathsurround\z@ \frozen@everymath{\mathgroup\macc@group\relax}%
  \macc@set@skewchar\relax
  \let\mathaccentV\macc@nested@a
  \macc@nested@a\relax111{#1}%
  \endgroup
}
\makeatother

\graphicspath{{figures/}}

\title[Bright vs faint LAEs]{Lyman-$\alpha$ emitters gone
    missing: the different evolution of the bright and faint populations}
    
\author[Weinberger et al.]{Lewis H. Weinberger$^{1}$\thanks{Email: 
\href{mailto:lewis.weinberger@ast.cam.ac.uk}{lewis.weinberger@ast.cam.ac.uk}},
Girish Kulkarni$^{1}$,
Martin G. Haehnelt$^{1}$,
\newauthor
Tirthankar Roy Choudhury$^{2}$ and
Ewald Puchwein$^{1}$
\\
$^{1}$ Institute of Astronomy and Kavli Institute for Cosmology, University
of Cambridge, Madingley Road, Cambridge CB3 0HA, UK
\\
$^{2}$ National Centre for Radio Astrophysics, Tata Institute of Fundamental 
Research, Pune 411007, India
}

\date{Accepted XXX\@. Received YYY; in original form ZZZ}

\pubyear{2018}

\begin{document}
\label{firstpage}
\pagerange{\pageref{firstpage}--\pageref{lastpage}}
\maketitle

\begin{abstract}
We model the transmission of the Lyman-$\alpha$ line through the 
circum- and intergalactic media around dark matter haloes expected to host 
Lyman-alpha emitters (LAEs) at $z \geq 5.7$, using the high-dynamic-range 
Sherwood simulations.  
We find very different CGM environments around more massive 
haloes ($\sim 10^{11} \mathrm{M}_\odot$) compared to less massive haloes 
($\sim 10^{9} \mathrm{M}_\odot$) at these redshifts, which can contribute to
a different evolution of the Ly$\alpha$ transmission from LAEs within these 
haloes.
Additionally we confirm that part of the
differential evolution could result from bright LAEs being more likely to
reside in larger ionized regions.
We conclude that a combination of the CGM environment and the IGM
ionization structure is likely to be responsible for the 
differential evolution of the bright and faint ends of the LAE luminosity 
function at $z \geq 6$.
More generally, we confirm the suggestion that the self-shielded neutral gas in
the outskirts of the host halo can strongly attenuate the Ly$\alpha$
emission from high redshift galaxies. We find that this has a stronger effect on
the more massive haloes hosting brighter LAEs.
The faint-end of the LAE luminosity function is thus a 
more reliable probe of the average ionization state of the IGM\@.
Comparing our model for LAEs with  a range of observational data we find that 
the favoured reionization histories are our previously advocated `Late' and 
`Very Late' reionization histories, in which reionization  finishes rather 
rapidly at around $z\simeq6$.
\end{abstract}



\section{Introduction}
\label{sec:introduction}
Observations such as the Lyman-$\alpha$ (Ly$\alpha$)
forest in quasar spectra \citep[][]{2006AJ....132..117F,
2015MNRAS.447.3402B,2015MNRAS.447..499M} and the Thomson optical 
depth to the CMB \citep[][]{2016A&A...596A.108P} suggest
that the neutral hydrogen fraction of the intergalactic medium (IGM) increases 
between 
redshifts of $z\sim6$ and $z\sim10$, during the Epoch of Reionization (EoR).
This final phase transition of the Universe is, however, not yet completely understood;
in particular there is still some debate about the contribution of different 
sources responsible for the reionization of hydrogen \citep{2015MNRAS.451.2030D}. 
To make progress requires further improved knowledge of the luminosity functions
and the escape fractions of ionizing photons for possible candidates, 
for which the faint end is particularly challenging at high redshifts
\citep{2016MNRAS.463.1968Y,2015ApJ...803...34B,2017arXiv171009390M,
2006AJ....131.2766R}. Firmer constraints on the exact redshifts at which the
reionization process began and ended are also challenging to obtain, due to 
the still rather scarce data and the model-dependence of the constraints 
obtained from observations 
\citep[for a review of IGM models, see][and references therein]
{2009CSci...97..841C,2009RvMP...81.1405M}. \citet{2017MNRAS.465.4838G} for 
example used a Bayesian framework to combine a selection of observational 
results but noted, as other authors have, that there are degeneracies between 
the EoR parameters which cannot yet be broken by current observations.

One notable observation made in recent years is the
dramatic decline in the space density of Ly$\alpha$ emitting galaxies (LAEs)
beyond $z>6$ 
\citep{2006ApJ...648....7K,2010ApJ...723..869O,2010ApJ...725..394H,
2014ApJ...797...16K}, compared to continuum
selected galaxies \citep{2015ApJ...803...34B,2011ApJ...728L...2S,
2014ApJ...793..113P,2012ApJ...744..179S}.
Note that at lower redshifts ($3\lesssim z \lesssim 5$, after hydrogen reionization),
however, the LAE luminosity function shows little evolution
\citep{1998ApJ...502L..99H,2008ApJS..176..301O}.
With an increasingly neutral fraction of hydrogen beyond $z\sim6$, we expect 
more of the Ly$\alpha$ emission to be
absorbed and scattered by the IGM, and hence a reduction in observed flux 
compared to the continuum.
This has been used to obtain model-dependent constraints on the evolution of the 
neutral hydrogen fraction. For example \citet{2017arXiv170302501O} used the model
of \citet{2004MNRAS.349.1137S} to convert a Ly$\alpha$ transmission
ratio into a fraction $x^{z=7}_{\mathrm{HI}}\gtrsim$ 0.3 -- 0.4. 

There have been a number of analytic and numerical models developed to explain
the apparent rapid decline of Ly$\alpha$ emission from galaxies;
for example taking into account the role of dust and 
reionization \citep{2009MNRAS.400.2000D}, of self-shielded absorbers 
\citep{2013MNRAS.429.1695B,2015MNRAS.452..261C}, the infall of the CGM onto 
the host haloes \citep{2017ApJ...839...44S},
or ruling out the role of IGM attenuation as a sole
factor \citep{2015MNRAS.446..566M}. 

One of the difficulties in explaining
this decline is the dependence of the IGM transmission on the Ly$\alpha$ 
emission line profile of the galaxy, which is complicated by the Ly$\alpha$ 
radiative transfer out of the galaxy's interstellar medium (ISM). It has been 
found empirically
that the peak of the emission profile is often offset redwards from the Ly$\alpha$ 
frequency \citep{2010ApJ...719.1168E}. Studies at lower redshifts
have found correlations between this offset and emission properties such as 
line magnitude or equivalent width \citep{2016ApJ...820..130Y}.
For high redshifts the usual reference lines for determining
this offset (such as [\ion{O}{III}] or H$\alpha$) are not observable with 
ground based telescopes. This leaves either using scaling relationships from 
low redshift observations \citep{2014ApJ...795...33E} or,
if available, using detections of lines such as \ion{C}{III]}$\lambda1909$ 
\citep{2015MNRAS.450.1846S}.

Theoretical modelling of the Ly$\alpha$ emission profile
is made difficult by the resonant nature of the line, resulting in emission 
profiles that are strongly affected by the ISM 
\citep{2010ApJ...716..574Z}. Use of Monte Carlo radiative transfer 
codes \citep{2012MNRAS.425...87O,2016ApJ...826...14G} 
and analytic methods \citep{2006ApJ...649...14D} has led to simple 
parameterised models of the emission profile 
such as the shell model \citep{2015ApJ...812..123G}, but see 
for example \citet{2011MNRAS.416.1723B} for more realistic models.
The sensitivity of the emission profiles to the physical and dynamical state
of hydrogen in and around galaxies makes isolating the intrinsic galaxy
evolution from the IGM evolution very difficult.

Recent surveys probing beyond $z=7$ have found a further complication: some 
observers have measured a luminosity dependence for the attenuation of 
quantities such as the luminosity function and the LAE fraction 
\citep{2012MNRAS.422.1425C,2014ApJ...797...16K,
2017arXiv170302985Z,2015MNRAS.451..400M,2016MNRAS.463.1678S}.
Faint ($M_\mathrm{UV} > -20.25$) LAEs are observed to decline in
number in a similar manner beyond $z=7$ as was seen for $z=$ 5 --- 6, and this
has been
used to extrapolate reionization histories. For bright 
($M_\mathrm{UV} < -20.25$) LAEs however, a much
slower evolution has been observed. This can be most clearly seen in the 
luminosity functions of \citet{2017arXiv170302985Z} and 
\citet{2017arXiv170302501O}, as well as
the estimated LAE fraction in \citet{2017MNRAS.464..469S}. One suggested
explanation \citep[][]{2015ApJ...810L..12Z} for
this much weaker decline in the number of bright LAEs is that such galaxies
sit in (and contribute ionizing photons to) larger ionized bubbles, 
and hence are preferentially more visible than fainter galaxies.

There has been some recent theoretical work using simulations to explore the 
causes of these observations. \citet{2018arXiv180101891M} explored the effect of a
mass-dependent intrinsic velocity offset in the emission profile of LAEs,
finding that larger velocity offsets can increase the visibility of bright LAEs.
\citet{2018arXiv180100067I} explored the effect of a mass-dependent optical
depth in the host halo, and found such a dependence was required to explain
observations. In this work we will further explore such effects, as well as the
different roles the larger IGM environment can play around bright and faint LAEs.

There has also been some discussion in the literature of the effects of different
selection techniques used for characterising LAEs \citep{2010MNRAS.408.1628S}, 
which can be divided into two categories: (i) (broadband) UV-selection with spectroscopic
follow up \citep[as in][for example]{2011ApJ...728L...2S}, 
and (ii) direct (narrowband) Ly$\alpha$ selection 
\citep[as in][for example]{2014ApJ...797...16K}. 
We note that observed LAE fractions are found via the former method, whilst most
LAE luminosity functions are presented for populations found using the latter
technique. In both cases the selection effects 
(such as AGN contamination) may play an important role in the inferred
properties of high redshift LAEs. Importantly for our
modelling, the selection technique will affect the mapping between galaxy mass
and Ly$\alpha$ (or UV) luminosity. We discuss this further in section 
\ref{sec:selection}.

In this paper we use a semi-analytic treatment of reionization,
combined with the high-dynamic-range Sherwood simulations \citep{2017MNRAS.464..897B},
to explore the effect of the IGM environment on the luminosity-dependent LAE evolution. 
In section \ref{sec:method} we outline our simulation
setup and calibration, which is based on \citet{2015MNRAS.452..261C}. 
Section \ref{sec:calculations} describes the
framework we employ for calculating the transmission of Ly$\alpha$ radiation 
through the IGM\@. We establish models for reionization and for the LAEs in section
\ref{sec:Models}. We then present our results for these different models 
in section \ref{sec:results}. In section \ref{sec:discussion} we discuss these 
results in comparison to other work, and finally draw conclusions in 
section \ref{sec:conclusion}.


\section{Simulation Method}
\label{sec:method}
In order to investigate the role of the IGM on LAE observations, we use  
cosmological hydrodynamical simulations with a semi-analytic treatment of
reionization. There are two components to our numerical modelling:
(i) a simulation of the (partially reionized) IGM, which includes the spatial
distribution of neutral hydrogen, the peculiar velocities of the IGM gas and 
its temperature; (ii) a source model that produces galactic Ly$\alpha$ 
emission, which accounts for
the spatial distribution of LAEs and their emission profiles. 

For step (i), the simulation of the IGM, we follow the procedure in 
\citet{2015MNRAS.452..261C}, hereafter referred to as CPBH15.
This approach starts from a cosmological hydrodynamic
simulation, and then applies the excursion set formalism 
\citep{2004ApJ...613....1F,2011MNRAS.411..955M,2007ApJ...654...12Z} to determine the
large-scale ionization structure. We apply a 
self-shielding prescription that models the occurrence of neutral hydrogen embedded in
ionized regions. Our reionization simulations are then calibrated to three different
reionization histories, spanning the range consistent with CMB and Ly$\alpha$
forest data. This first step of the simulation is outlined in sections
\ref{sec:LSIS} through \ref{sec:calib}.
For step (ii), the source model, we also start with the same basic model used 
in CPBH15, aiming to reproduce their results. We then extend this basic model 
to try to account for differences between bright and faint LAEs. The details 
of these source models are outlined in section \ref{sec:sources}.

\begin{figure*}
    \includegraphics[width=2\columnwidth]{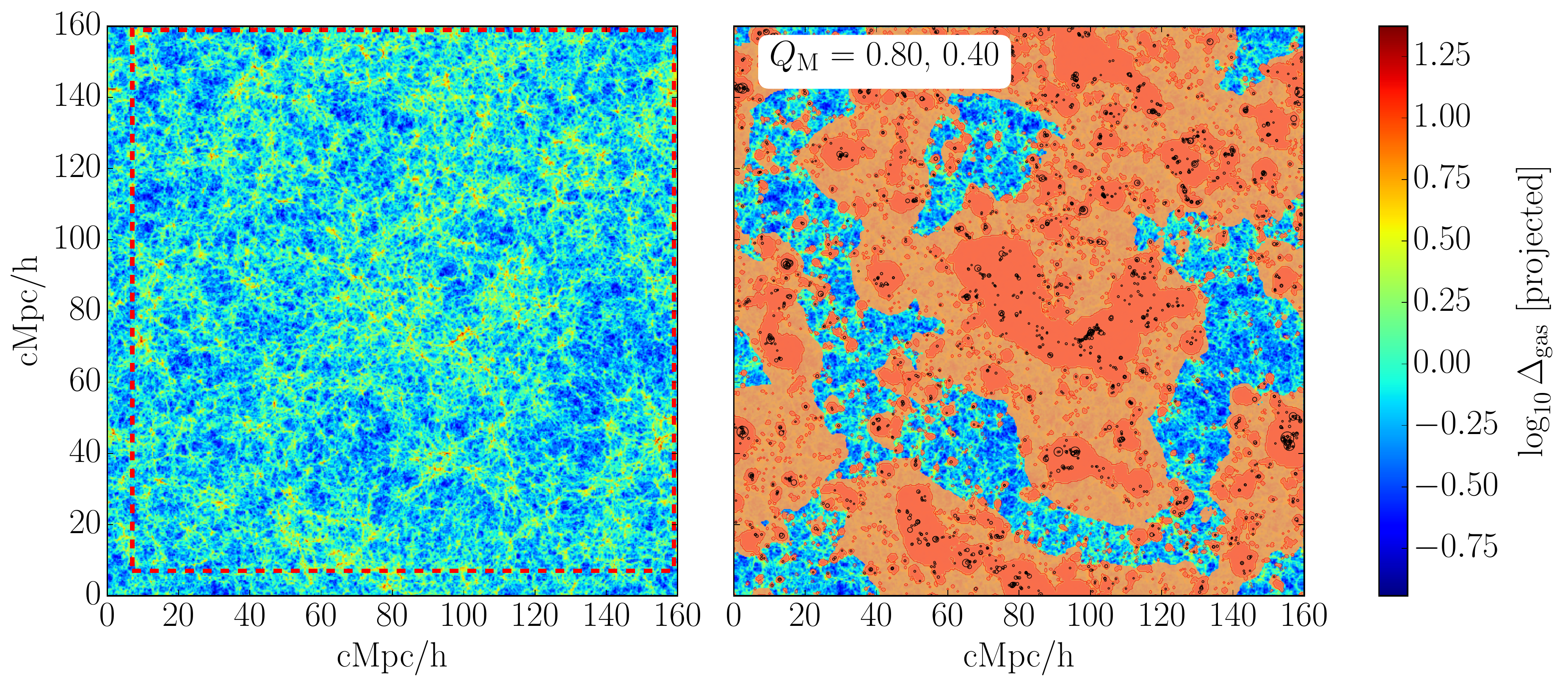}
    \caption{Gas overdensity (left) and ionization field (right, shaded) in a 
        projected 0.5 cMpc/h slice of the $L=160$, $N=2048^3$, $z=7$ snapshot from
        the Sherwood suite. The density has been interpolated onto a uniform 
        3D grid using the SPH kernel, and then a thin slice has been
        projected to create the 2D map. The dashed red square is a 
        representative area for the comoving volume surveyed by 
        \citet{2017arXiv170501222K}. For the ionization structure on the 
        right panel, the mass-averaged ionized fraction has been
        set to values of $Q_\mathrm{M} = 0.8$ and $0.4$. The shaded areas mark 
        ionized regions (found using the excursion set method) with the 
        darker (brighter) shade showing the $Q_\mathrm{M} = 0.4$ ($0.8$) case.
        The positions of haloes from this slice are shown in black, with marker
        size proportional to halo mass. 
    }
    \label{fig:bubble}
\end{figure*}

\subsection{Large-scale ionization structure}
\label{sec:LSIS}
The underlying cosmological hydrodynamical simulation used in this work is 
taken from the 
Sherwood simulation suite \citep{2017MNRAS.464..897B}, initially run as part 
of a \textsc{PRACE} simulation program. These simulations were run using
a modified version of the \textsc{P-Gadget-3} TreePM 
smoothed-particle-hydrodynamics (SPH) code, itself an updated version of 
\textsc{Gadget-2} 
\citep{doi:10.1111/j.1365-2966.2005.09655.x,2001NewA....6...79S}. Dark matter
haloes were found on the fly using a Friends-of-Friends algorithm, with a minimum
particle number of 32. The $\Lambda$CDM 
parameters used for this run (and hereafter in this work) are based on the 
\citet{2014A&A...571A..16P}
results: $h=0.678$, $\Omega_m = 0.308$, $\Omega_\Lambda = 0.692$, 
$\Omega_b = 0.0482$, $\sigma_8 = 0.829$, $n = 0.961$, and 
$Y_{\mathrm{He}} = 0.24$. The simulation
used for this work was performed in a box of length $L=160$ 
$\mathrm{cMpc/h}$\footnote{For test cases to compare with CPBH15
we also used a box of length $L=80 \:\: \mathrm{cMpc/h}$.} (where prefix c
indicates comoving, and prefix p indicates proper). The runs were started
with $2048^3$ particles of gas and dark matter each ($N=2\times2048^3$ 
total), giving a dark matter mass
resolution of $M_{\mathrm{DM}}=3.44\times10^7$ $\mathrm{M_{\odot}/h}$.
The gravitational softening length was set at 
$l_{\mathrm{soft}}=3.13$ $\mathrm{ckpc/h}$.
Snapshots of the initial \textsc{PRACE} run were saved for redshifts
in the epoch of interest at $z = 6.0$, 7.0, 8.0 and 10.0. We have also re-run
the simulation in order to better sample the EoR, saving snapshots every 40 Myrs;
in particular in this work we make use of snapshots at $5.756 \leq z \leq 9.546$.

Alongside the hydrodynamical and gravitational evolution of the gas and 
dark matter, the simulation
included photo-ionization and photo-heating calculated using the spatially
uniform background from \citet{2012ApJ...746..125H}.\footnote{At lower redshifts
not considered here, the photo-heating rates were somewhat modified to better 
match the temperature measurements of \citet{2011MNRAS.410.1096B}.}

Note that for these simulations, the \texttt{QUICK$\_$LYALPHA} star formation
implementation in \textsc{P-Gadget-3} was used.
This option speeds up the simulation by converting gas particles (with 
temperatures lower than $10^5 \:K$ and 
overdensities larger than a thousand times the mean baryonic density) into 
collisionless star particles 
\citep{2004MNRAS.354..684V}. For use in calculations
the densities, velocities and temperatures of the particles were projected 
onto a grid using the SPH kernel. 
A projected slice of the density field from the simulation at redshift $z=7$ 
can be seen in the left panel of Figure \ref{fig:bubble}.

In CPBH15 a hybrid simulation was employed, comprised of an $L=10$ cMpc/h, 
$N=2\times512^3$ \textsc{P-Gadget-3} hydrodynamical simulation to model the
hydrogen distribution, and a larger low resolution dark matter-only simulation 
with $L=100$ cMpc/h, $N=1200^3$. These simulations were combined by
tiling the small simulation box across the larger volume, making use of the 
ionization structure and large-scale velocity modes of the  large simulation box.
We take advantage here of the much higher dynamic range of the 
Sherwood simulation suite and employ  instead a single hydrodynamical 
simulation with almost twice the volume of their hybrid simulation, but at a
factor two  reduced spatial resolution compared to their 10 cMpc/h sized 
hydrodynamical simulation. Although lower in resolution this has the advantage
of retaining the
correlation between the gas density fields and the halo structure of the 
simulation, which was not present in the hybrid simulation of CPBH15. The 
larger volume also allows us to probe to higher halo masses, which is key to 
our modelling of bright and faint LAEs.

In recent observations, \citet{2017arXiv170501222K} 
surveyed comoving volumes of $\sim1.2\times10^{7}$ Mpc$^{3}$; our simulation 
volume ($\sim 1.3\times10^{7}$ Mpc$^{3}$) is therefore a better representation than the 
smaller volume of CPBH15 ($\sim 0.3\times10^{7}$ Mpc$^{3}$). In Figure \ref{fig:bubble}
we show a representative survey area with a red dashed square for comparison 
with our box size. 

To generate the large-scale ionization structure of the simulation, we apply 
an excursion set method
\citep{2004ApJ...613....1F,Mesinger:2007pd,2009MNRAS.394..960C,
2011MNRAS.411..955M, 2010MNRAS.406.2421S, 2016MNRAS.457.1550H}.
This is a semi-analytic approach, which has been found to reproduce ionization
fields that agree with low-resolution radiative transfer simulations 
\citep[e.g. in][]{2014MNRAS.443.2843M}, whilst being computationally 
efficient. The first step assigns to haloes an emissivity as a function of
their mass. In this work we assume a linear relationship, with the number of ionizing
photons produced by a halo, $N_\gamma=c_\gamma M_{\mathrm{h}}$, with $c_\gamma$
a constant of proportionality. Although recent observations of high-redshift UV
luminosity functions may suggest non-linear scalings \citep[e.g.][]{2015ApJ...813...21M},
this simplifying assumption should not have a strong effect on our modelling. 
In earlier work such a linear scaling for the ionizing photon budget
with mass has been found to approximately reproduce the high-redshift UV luminosity functions
 \citep{2010ApJ...714L.202T}. We note that non-linear models
were used by \citet{2016MNRAS.463.2583K} in the same reionization framework used
here, but were not found to have a strong effect on the resulting reionization 
history. Furthermore, \citet{2015MNRAS.453.2943C} employed full radiative transfer simulations,
modelling the ionizing luminosity with a similar linear scaling, and they were 
able to reproduce observations of the end of reionization very well.
Physically, this simplifying assumption of a linear scaling may break down if
galactic outflows or feedback alter the dependence \citep{2011MNRAS.410.1703F}.
We do not impose a minimum mass of star forming
haloes, but use the entire halo population of the simulation. Note that, 
at least initially the value of $c_\gamma$ is not important because of
the later calibration scheme (see section \ref{sec:calib}).
Using this relationship we establish a radiation field based on the locations 
and masses of the haloes in the simulation. We then flag a cell in the box,
$\mathbf{i}$, as ionized if there is some radius $R$ inside which the condition,
\begin{equation}
    \label{eq:excursion1}
    \langle n_\gamma (\mathbf{i}) \rangle_R > \langle 
    n_{\mathrm{H}} (\mathbf{i}) \rangle_R \: (1 + \widebar{N}_{\mathrm{rec}}),
\end{equation}

\noindent is satisfied. Here $n_\gamma$ and $n_{\mathrm{H}}$ are the photon 
and hydrogen (comoving) number densities, respectively. The averages are taken 
over a spherical region of radius $R$ centred on the cell. This condition is 
therefore comparing the number of ionizing photons in the neighbourhood of the
cell (at a given scale $R$) to the number of hydrogen atoms in the same region, and 
if it is larger for some value of $R$ then we flag that cell as being ionized. 
The factor of $1 + \widebar{N}_{\mathrm{rec}}$ accounts for recombinations, where
$\widebar{N}_{\mathrm{rec}}$ is the average number of recombinations per 
hydrogen atom that occur in the IGM\@. An equivalent statement of the condition 
in Eq.~(\ref{eq:excursion1}) is that a cell will be ionized if
\citep{2009MNRAS.394..960C},
\begin{equation}
    \label{eq:excursion2}
    \zeta_{\mathrm{eff}} f_{\mathrm{coll}}(\mathbf{i},R) \geq 1,
\end{equation}

\noindent where $\zeta_{\mathrm{eff}} = 
c_\gamma m_H\left((1+\widebar{N}_{\mathrm{rec}})(1- Y_{\mathrm{He}})\right)^{-1}$
is an efficiency parameter and $f_{\mathrm{coll}}$
is the collapsed mass within a spherical volume of radius $R$ given by,
\begin{equation}
    \label{eq:fcoll}
    f_{\mathrm{coll}}(R) = \frac{1}{\widebar{\rho}(R)} 
    \int_{M_\mathrm{min}}^\infty dM \left. \frac{dn}{dM}\right|_R\: M.
\end{equation}

\noindent In Eq.~(\ref{eq:fcoll}), $\widebar{\rho}(R)$ is the average matter 
density within a radius $R$. Note that to go from Eq.~(\ref{eq:excursion1}) to 
Eq.~(\ref{eq:fcoll}) we have used the linear relationship for $N_\gamma (M)$.
The constant of proportionality, $c_\gamma$, and the 
recombination factor have been absorbed into $\zeta_{\mathrm{eff}}$. This 
ionization efficiency parameter controls the size of ionized bubbles, and
must be calibrated so that the mass-averaged
neutral fraction, $Q_\mathrm{M}$ matches the desired reionization history. 
Using the above prescription we can determine for each cell in the simulation 
whether it is ionized. For those cells
that are not ionized we set the neutral fraction to $x_\mathrm{HI} = 1$. If a 
cell is ionized, then its neutral fraction is found assuming photoionization 
equilibrium \citep{2009RvMP...81.1405M},
\begin{equation}
    \label{eq:xeq}
    x_\mathrm{HI}(\mathbf{i}) =  
    \frac{n_e(\mathbf{i})\: \alpha_\mathrm{B}(T)}{\Gamma_\mathrm{HI} 
    + n_e(\mathbf{i})\: \alpha_\mathrm{B}(T)},
\end{equation}

\noindent where $n_e$ is the free electron number density, 
$\alpha_\mathrm{B}(T)$ is the case B recombination rate and 
$\Gamma_\mathrm{HI}$ is the background photoionization rate. Note that here
we make the simplifying assumption of  a spatially uniform background photoionization 
rate (within ionized regions). In reality, however, the varying position of sources and the 
inhomogeneous distribution of matter should lead to a non-uniform background.
To include this effect properly would require full radiative transfer calculations.
We explore how varying the photoionization rate can affect LAE visibility in
section \ref{sec:sadoun}. This suggests that an inhomogeneous UV background could lead 
to fluctuations in the Ly$\alpha$ transmission from halo to halo, 
but it is expected that this would be subdominant
to the average evolution driven by the IGM.
The values of $Q_\mathrm{M}$ and $\Gamma_\mathrm{HI}$ are found during
the calibration stage, such that the simulation is consistent with a desired 
reionization history. 

In the right hand panel of Figure
\ref{fig:bubble} we show the ionization field 
for two different ionized fractions: in dark orange we show the ionized regions
for $Q_\mathrm{M}=0.4$, whilst the lighter orange region is at a higher 
fraction of $Q_\mathrm{M}=0.8$. The positions of the haloes are overplotted as
empty black circles, with the size of the marker proportional to the halo mass.
As expected from the excursion set construction, the largest haloes sit in and
dominate the largest ionized regions.

\subsection{The self-shielding of dense gas in ionized regions}
\label{sec:SS}
The excursion set method described above is effective at producing the 
large-scale ionization field, however one of its most significant limitations 
is modelling dense self-shielded clumps within already (re-)ionized regions.
In order to account for such regions, we employ a prescription based on the 
overdensity of hydrogen, $\Delta_\mathrm{H}$, in the hydrodynamical 
simulation.  The prescription we use is based on the results of 
\citet{2017arXiv170706993C},  a modified version of   those 
found in \citet{2013MNRAS.430.2427R} that aims to reproduce the   self-shielding 
of dense regions within ionized bubbles during reionization.
Here we apply the same ionization equilibrium
approach from Eq.~(\ref{eq:xeq}), but the local photoionization rate is modified to the 
empirical fit of \citet{2013MNRAS.430.2427R},
\begin{multline}
    \frac{\Gamma_\mathrm{HI}(\mathbf{i})}{\Gamma_\mathrm{HI,\:global}} = (1 - f(z)) 
    \left[ 1 + \left( \frac{\Delta_\mathrm{H}}{\Delta_\mathrm{ss}}
    \right)^{\beta(z)}\right]^{\alpha_1(z)}\\
    + f(z) \left[ 1 + \left( \frac{\Delta_\mathrm{H}}{\Delta_\mathrm{ss}} \right) \right]
    ^{\alpha_2(z)},
\end{multline}
where $\Delta_\mathrm{ss}$ is the overdensity threshold for self-shielding,
and $f$, $\beta$, $\alpha_1$, $\alpha_2$ are the redshift dependent
parameters found by \citet{2017arXiv170706993C}. We use the threshold found
by \citet{2017arXiv170706993C}, scaled appropriately by photoionization rate.
We note that they found the self-shielding threshold is in reasonable agreement
with the parametrization found by considering the local Jeans length
\citep{2001ApJ...559..507S,2005ApJ...622....7F}, 
\begin{multline}
    \label{eq:jeans_ss}
    \Delta_\mathrm{ss} = 36 \left( \frac{\Gamma_\mathrm{HI}}{10^{-12} \mathrm{s}^{-1}} \right)^{2/3}
    \left( \frac{T}{10^{4} \mathrm{K}} \right)^{2/15} \left( \frac{\mu}{0.61} \right)^{1/3}\\
    \times  \left( \frac{f_\mathrm{e}}{1.08} \right)^{-2/3} \left( \frac{1 + z}{8} \right)^{-3},
\end{multline}
where $\mu$ is the mean molecular weight. Our default prescription
does have a higher self-shielding threshold than was found for the `SS-R' case 
of CPBH15, in which they followed the \citet{2013MNRAS.430.2427R} prescription.
This means self-shielding plays a less dominant role in our models. We explore
the effect of changing this prescription in section \ref{sec:sadoun}.

\subsection{Calibrating the simulations for different reionization histories}
\label{sec:calib}

The methodology described above creates a realistic large-scale ionization 
field, as well as an accurate neutral hydrogen distribution within ionized 
bubbles. The model has two free parameters however, $\Gamma_\mathrm{HI}$ and 
$Q_\mathrm{M}$, which need to be calibrated so that the simulation matches 
observational constraints. In order to calibrate these two 
quantities self-consistently, we iteratively solve the 
equation \citep{2016MNRAS.463.2583K,2009CSci...97..841C},
\begin{equation}
    \label{eq:grt}
    \frac{dQ_\mathrm{M}}{dt} = \frac{
    \langle \dot{n}_\mathrm{ion}\rangle - \langle \dot{n}_\mathrm{rec} \rangle}
    {n_{\mathrm{H}}},
\end{equation}
where $Q_\mathrm{M}$ is the \emph{mass-averaged} neutral fraction within the 
simulation box. We solve this equation by starting with the desired 
$Q_\mathrm{M}$ and a guessed $\Gamma_\mathrm{HI}$. This allows us to estimate 
the comoving emissivity \citep{2012MNRAS.423..862K,2013MNRAS.436.1023B},
\begin{equation}
    \langle \dot{n}_\mathrm{ion} \rangle = 
    \frac{\Gamma_\mathrm{HI}Q_\mathrm{V}}
    {(1+z)^2 \sigma_\mathrm{H} \lambda_\mathrm{mfp}}
    \left( \frac{\alpha_b + 3}{\alpha_s} \right) ,
    \label{eq:emi}
\end{equation}
where $Q_\mathrm{V}$ is the \emph{volume-averaged} neutral fraction, 
$\sigma_\mathrm{H}$ is the hydrogen photoionization cross-section at 912 \AA, 
$\lambda_\mathrm{mfp}$ is the mean free path of ionizing radiation at the same wavelength, and the 
bracketed factor includes the spectral indices for ionizing sources 
$\alpha_s$ and the ionizing background $\alpha_b$.
Note that in the simulation we calculate the mass and volume averaged neutral 
fractions by summing over the neutral fraction in each projected grid cell, 
$q(\mathbf{i})$, with the appropriate weighting,
\begin{align}
\begin{split}
    \label{eq:Q_M}
    Q_\mathrm{M} ={}& \frac{1}{M_\mathrm{tot}} \sum_\mathbf{i} 
    M(\mathbf{i}) q(\mathbf{i}) = \frac{1}{\rho_\mathrm{tot}} \sum_\mathbf{i} 
    \rho(\mathbf{i}) q(\mathbf{i}) ,
\end{split}\\
\begin{split}
    \label{eq:Q_V}
    Q_\mathrm{V} ={}& \frac{1}{N} \sum_\mathbf{i} q(\mathbf{i}) ,
\end{split}
\end{align}
where $M(\mathbf{i})$, $\rho(\mathbf{i})$ are the mass and density in a given cell, 
the total mass and density are $M_\mathrm{tot} = \sum_\mathbf{i} M(\mathbf{i})$
and $\rho_\mathrm{tot} = \sum_\mathbf{i} \rho(\mathbf{i})$ respectively,
and $N$ is the total number of cells (e.g $2048^3$).
These expressions are valid here because of the uniform grid projection.

The mean free path is fixed to the predicted values of a given reionization 
history model (see section \ref{sec:Models}). 
We found this to be more stable than trying to  calculate the mean free path 
iteratively from the  simulation using Eq.~(\ref{eq:grt}). To test that this was not 
sensitive  to the resolution needed to properly resolve the self-shielded regions 
such as DLAs and LLSs, we calculated the mean free path from the simulation
for a fixed photoionization rate. Our  calculations are indeed converged with respect 
to the predicted values from the models. This suggests that although we have to 
fix the mean free path for the calibration, we do properly resolve the 
self-shielded systems.

The bracketed term on the right of Eq.~(\ref{eq:emi}) is determined by the 
spectrum of ionizing radiation; in this work we use the same value as used 
by \citet{2012ApJ...746..125H}. During the iterative solving of Eq.~(\ref{eq:grt})
we also find the globally averaged comoving rate of recombinations, given by,
\begin{align}
\begin{split}
    \langle \dot{n}_\mathrm{rec} \rangle ={}& \frac{1}{N} 
    \sum_{\mathbf{i}} \alpha_B (1+z)^3
    \: {n}_\mathrm{e}(\mathbf{i}) \: {n}_\mathrm{HII}(\mathbf{i})
\end{split}\\
\begin{split}
    \simeq{}& \frac{1}{N} \sum_{\mathbf{i}} \alpha_B (1 + z)^3 \: 
    f_\mathrm{e} \: {n}_\mathrm{HII}^2(\mathbf{i}),
\end{split}
\end{align}
where $\alpha_B$ is the (case-B) recombination rate, and $f_\mathrm{e}=1.08$ 
is the number of electrons per hydrogen nucleus\footnote{
Note $f_\mathrm{e}>1$ due to singly ionized Helium in the HII regions.}.

In summary, the calibration method takes as input the values for $Q_M(z)$
and $\lambda_\mathrm{mfp}(z)$ from each reionization history model, and then
uses the large-scale ionization field (constructed via the excursion set method)
to solve for an equilibrium $\Gamma_\mathrm{HI}$ that satisfies Eq.~\ref{eq:grt}.


\section{Lyman-\texorpdfstring{$\alpha$}{[alpha]} Transmission}
\label{sec:calculations}

Having performed the calibration as detailed in section \ref{sec:calib}, we 
have simulation snapshots with realistic neutral hydrogen distributions that 
can be used to test the effect of the CGM and IGM on the transmission of the Ly$\alpha$ 
radiation from LAEs. 

Early galaxies with high star-formation rates (SFRs) produce ionizing radiation
in their stellar component \citep{1967ApJ...147..868P}. This ionizing radiation
is then converted into Ly$\alpha$ line emission through recombination and 
collisional excitation of the gas in the ISM \citep{1993ApJ...415..580C,
2014PASA...31...40D}. The radiative transfer of Ly$\alpha$ photons through the
ISM and CGM causes a diffusion in both physical and frequency space, resulting
in a significant change to the emission profile. The photons that escape the 
galaxy must then traverse the IGM, which at $z>6$ contains a significant 
non-zero neutral hydrogen fraction. Due to the resonant nature of Ly$\alpha$ 
absorption in neutral hydrogen, the presence of even small neutral fractions 
can alter the visibility of LAEs \citep[see][for reviews
of IGM and Ly$\alpha$ physics]{2009RvMP...81.1405M,2014PASA...31...40D}. 

As discussed in section \ref{sec:introduction}, observations of LAEs at high redshifts 
have found a decline in number densities. Explaining these observations is 
made difficult by the degeneracy between internal galaxy evolution 
(parameterized by the fraction of Ly$\alpha$ photons that escape galaxies,
$f_\mathrm{esc,Ly\alpha}$, which may be a function of $z$) and IGM absorption 
(parameterized by the neutral fraction $x_\mathrm{HI}$) 
\citep{2009MNRAS.400.2000D}. In this work we consider the effect of the
CGM/IGM only, and do not model galaxy evolution.

\subsection{Ly\texorpdfstring{$\alpha$}{[alpha]} transmission fraction}
\label{sec:define_T}
In order to quantify the effect of the IGM and CGM on the transmission of Ly$\alpha$ 
photons, we extract sightlines
from our simulation snapshots that pass through LAE host haloes, and 
calculate the radiative transfer along them (see section \ref{sec:Models} for
details on how LAE host haloes are selected). 
The sightlines are chosen to be 160 cMpc/h in length, parallel to the
simulation box axes. We take advantage of the periodic boundary conditions of
the simulation to translate the start of the sightline such that the halo is positioned at 
the centre\footnote{The optical depth calculation was found to converge on 
considerably smaller spatial scales than 80 cMpc/h.}. The gas properties are
initially gridded into 2048 bins (78.13 ckpc/h bin resolution), 
with a further 2048 bins in a high resolution
region of length 20 cMpc/h (giving 9.77 ckpc/h bin resolution) containing the 
host halo\footnote{We note that the softening length used in these simulations is
3.13 ckpc/h.}. This ensures we resolve
the gas around the host halo, including small-scale high density regions likely 
to self-shield. 

Neglecting scattering by dust, the equation of radiative transfer can be 
written \citep{2011piim.book.....D},
\begin{align}
    \begin{split}
        J_\nu(\tau_\nu) ={}& J_\nu(0) e^{-\tau_\nu} + \int^{\tau_\nu}_{0} 
        d\tau'_\nu \: e^{-(\tau_\nu - \tau'_\nu)}\: S_\nu(\tau'_\nu),
    \end{split}\\
    \begin{split}
        ={}& J_\nu(0) e^{-\tau_\nu},
    \end{split}
\end{align}
where $J(\nu)$ is the galaxy emission profile (the specific intensity of 
radiation at frequency $\nu$), $\tau(\nu)$ is the Ly$\alpha$ optical depth 
(see subsection \ref{sec:def_optdepth} below),
and the source function, $S_\nu$, is approximately zero because the Ly$\alpha$ 
emissivity of the IGM gas is negligible \citep{2016MNRAS.462.1961S}. 
This expression allows us to calculate the
emission profile of a galaxy after re-processing by the surrounding IGM gas,
$J'(\nu) = J(\nu) e^{-\tau(\nu)}$. With this we can calculate the transmission 
fraction of photons (or transmissivity) given by
\citep{2015MNRAS.446..566M},
\begin{equation}
    \label{eq:transmissivity}
    T^\mathrm{IGM}_\mathrm{Ly\alpha} = \frac
    {\int_{\nu_\mathrm{min}}^{\nu_\mathrm{max}} d\nu \: J(\nu) \: e^{-\tau(\nu)}}
    {\int_{\nu_\mathrm{min}}^{\nu_\mathrm{max}} d\nu \: J(\nu)},
\end{equation}
where $J(\nu)$ is appropriately normalized. Since we place the Ly$\alpha$ emitter
at the centre of the sightline, the frequency limits in 
Eq.~(\ref{eq:transmissivity}) used in this work are the Ly$\alpha$ frequency
blue/redshifted along half the sightline length, which extends considerably
beyond the wings of the emission profile.

\begin{figure*}
   	\includegraphics[width=2\columnwidth]{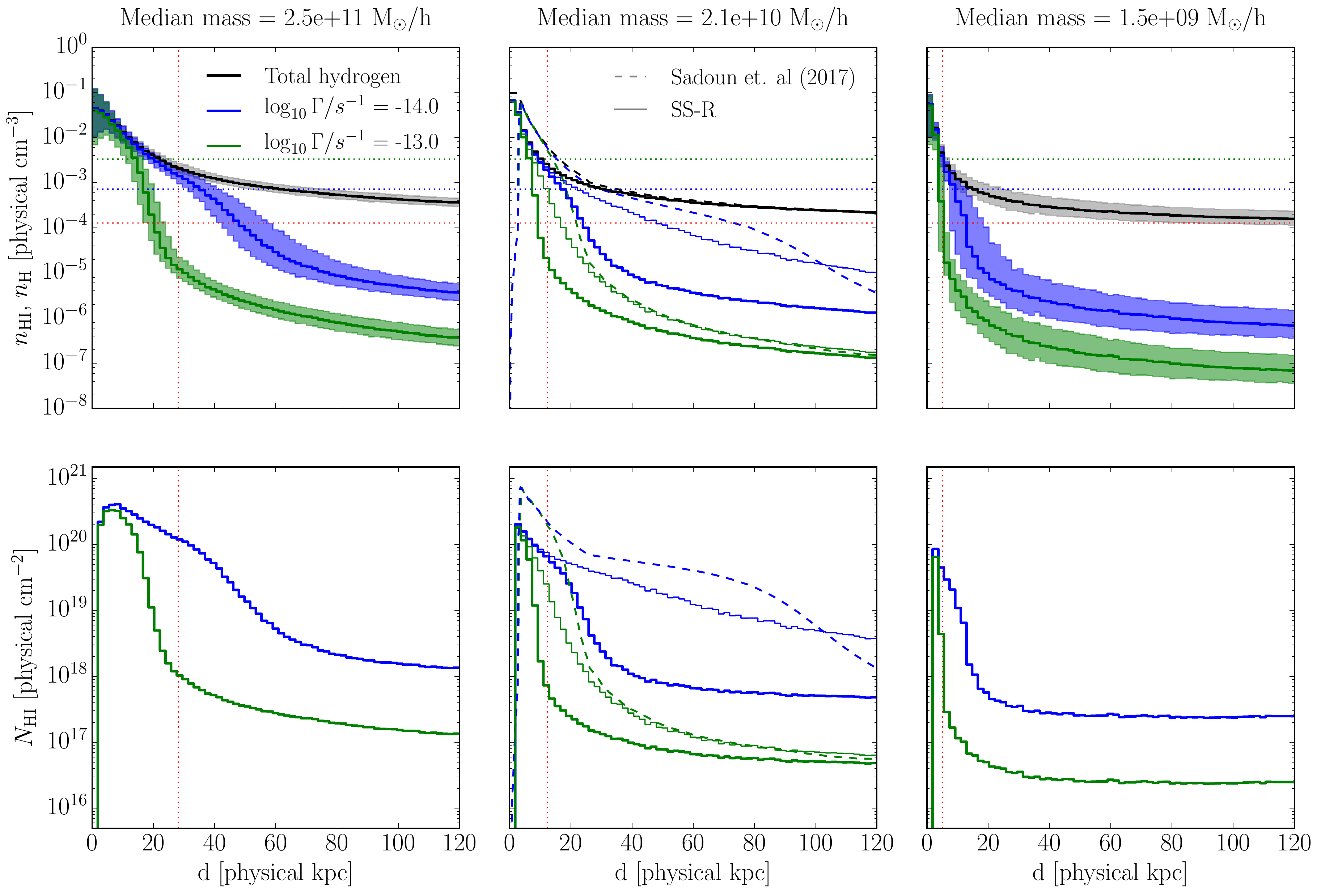}
    \caption{\emph{Top panels}: Median hydrogen number density profiles at $z=7$, 
    spherically averaged, for both total hydrogen in black, and neutral 
    hydrogen in green ($\Gamma = 10^{-13}$ s$^{-1}$)
    and blue ($\Gamma = 10^{-14}$ s$^{-1}$). The three panels correspond to
    samples of 500 haloes with mass bins that bracket the halo population of 
    our simulations; most massive on the left, least massive on the right, and
    a sample chosen for comparison with SZM17's $M_h = 10^{10.5}\:M_\odot$ model
    in the middle panel. The red vertical dotted line shows the median virial 
    radii, whilst the horizontal red line showns the mean baryonic density. 
    The horizontal green and blue dotted lines show the self-shielding density
    thresholds for the different photoionization rates. In
    the middle panel we also show the profiles from SZM17 with dashed lines, and
    the profiles found using the \citet{2013MNRAS.430.2427R} self-shielding 
    prescription (labelled SS-R) shown with thin stepped lines.
    The shaded regions indicate 68\% scatter in the samples (not shown in the
    middle panel to aid visual comparison with SZM17).
    \emph{Bottom panels}: Column densities at a given radius estimated by
    multiplying the number density by the radial distance.}
    \label{fig:n_HI_profiles}   
\end{figure*}

We note that this ``$e^{-\tau}$ modelling'' of the radiative transfer of Ly$\alpha$
photons has been compared to full radiative transfer by 
\citet{2010ApJ...716..574Z}. They suggested that
such models can over-attenuate the line profile compared to that of full
calculations because some of the frequency diffusion is neglected. 
A balance has to be struck between the frequency diffusion in the inner parts of the 
galaxy and the attenuation by the neutral hydrogen surrounding the galaxy. 
We will account for the frequency diffusion in the inner part of the galaxies in our 
modelling of the spectral distribution (see section \ref{sec:Models}).

\subsection{Ly\texorpdfstring{$\alpha$}{[alpha]} attenuation due to the CGM and IGM}
\label{sec:def_optdepth}
As suggested in \citet{2014PASA...31...40D} we  split  the Ly$\alpha$ 
optical depth responsible for attenuating the Ly$\alpha$ emission from galaxies into 
two contributions: 
(i) $\tau_\mathrm{HI}(z, v)$, the opacity due to any recombined 
neutral hydrogen or self-shielded regions \emph{within} ionized bubbles; 
(ii) $\tau_\mathrm{D}(z, v)$, the opacity due to damping-wing
absorption in the residual neutral IGM\@. Note that these quantities depend on 
the velocity offset, $v$, which is determined by both the Hubble flow and
the difference in peculiar velocity of emitter and absorber. 
So we can calculate,
\begin{equation}
    \label{eq:tau_decomp}
    \tau_\mathrm{Ly\alpha}(v) = \tau_\mathrm{HI}(v) + 
    \tau_\mathrm{D}(v).
\end{equation}
Physically, photons emitted close to line centre will redshift out of 
resonance as they traverse the IGM\@. It is important to consider that 
scattering/absorption occurs at velocity shifts close to zero in the 
absorber's rest frame. This means that redshifted photons in the frame of 
neutral gas infalling onto the host halo can be blueshifted back into resonance.

\begin{figure*}
   	\includegraphics[width=1.8\columnwidth]{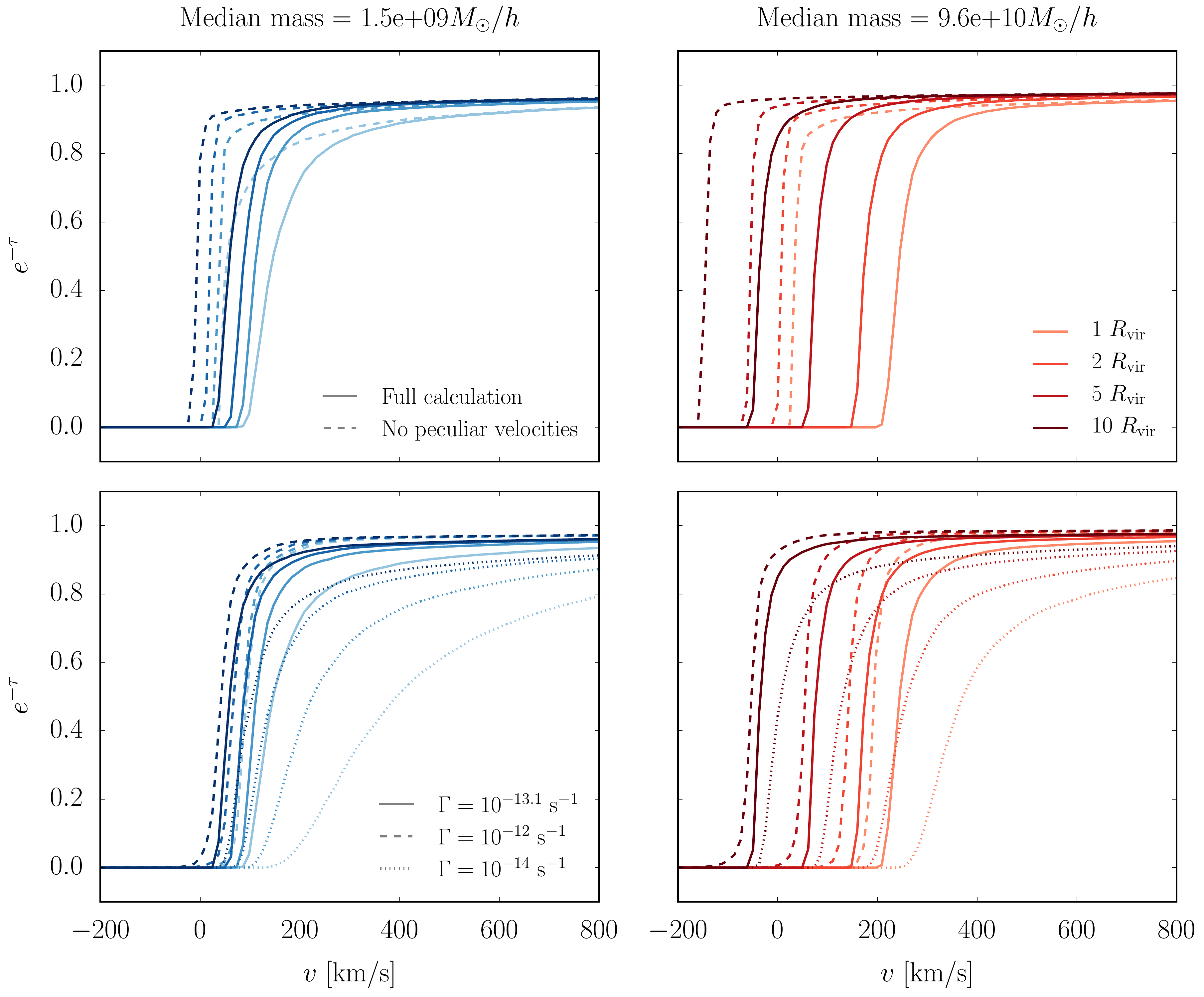}
    \caption{Median transmission curves, testing different exclusion radii
    (indicated by darkness of line), for a less massive sample of haloes
    (left panels) and a more massive sample (right panels). In the top panels the 
    solid lines show the results from the full calculation, whilst the dashed 
    lines show what happens when peculiar velocities are neglected. Both of
    these panels assume $\log_{10}\Gamma$/s$^{-1}$= -13.1.  The
    bottom panels compare three different photoionization rates: 
    $\log_{10}\Gamma$/s$^{-1}$= -12 (dashed), -13.1 (solid), -14 (dotted).
}
    \label{fig:sadoun_transmission}   
\end{figure*}

We note that both of the components in Eq.~(\ref{eq:tau_decomp}) are calculated 
in the same manner. In order to 
calculate the optical depth we assume a Voigt profile for the absorption cross section, in 
particular using the analytic approximation from \citet{2006MNRAS.369.2025T},
\begin{multline}
    \label{eq:Hjerting}
        H(a,x) = e^{-x^2} - \frac{a}{\sqrt{\pi}x^2}[e^{-2x^2} 
        (4x^4 + 7x^2 + 4 + 1.5x^{-2}) \\ - 1.5x^{-2} - 1],
\end{multline}
where $H(a,x)$ is the Hjerting function, related to the Voigt profile by 
\citep{1985rpa..book.....R},
\begin{equation}
    \label{eq:Voigt}
    \phi(\nu) = (\Delta \nu_\mathrm{D})^{-1} \pi^{-1/2}H(a,x),
\end{equation}
\begin{align}
\begin{split}
    \label{eq:dopp}
    \mathrm{where}\:\:\Delta \nu_\mathrm{D} \equiv{}& 
    \frac{\nu_\alpha}{c}\sqrt{\frac{2k_B T}{m_\mathrm{H}}},
\end{split} \\
\begin{split}
    \label{eq:aparam}
    a \equiv{}& \frac{\Lambda_\alpha}{4\pi\Delta \nu_\mathrm{D}},
\end{split} \\
\begin{split}
    \label{eq:xparam}
    x \equiv{}& \frac{\nu - \nu_\alpha}{\Delta \nu_\mathrm{D}}.
 \end{split}
\end{align}
Note in the above formulae we have used: the Ly$\alpha$ frequency 
$\nu_\alpha = 2.46\times10^{15}\mathrm{Hz}$, the hydrogen $2p\rightarrow1s$ 
decay rate $\Lambda_\alpha = 6.25\times10^8 \mathrm{s^{-1}}$, the Boltzman 
constant $k_B$, the hydrogen atomic mass $m_\mathrm{H}$ and the temperature of 
the gas, $T$, at the absorber.
For a given sightline, we find the optical depth in a (redshift-space) cell $\mathbf{i}$ by 
summing up all the contributions from 
positions in front of the emitter \citep{2007MNRAS.374..493B}, 
where we define $v=0$ at the position of the emitter,
\begin{equation}
    \label{eq:tau}
    \tau(\mathbf{i}) = \frac{\nu_\alpha \sigma_\alpha \delta R}{\sqrt{\pi}} 
    \sum_j^N \frac{n_\mathrm{HI}(j)}{\Delta \nu_\mathrm{D}(j)}H(a,x(i,j)),
\end{equation}
where $\delta R$ is the cell width, and the cell has Hubble velocity 
$v_\mathrm{H}(\mathbf{i})$ and peculiar velocity $v_\mathrm{pec}(\mathbf{i})$, 
such that
\begin{equation}
    \label{eq:xparam2}
    x(i,j) = \sqrt{\frac{m_\mathrm{H}}{2k_B T}}
    \:[ v_\mathrm{H}(\mathbf{i}) - v_\mathrm{H}(\mathbf{j})
    - v_\mathrm{pec}(\mathbf{j})].
\end{equation}
Eq.~(\ref{eq:tau}) is the optical depth to Ly$\alpha$ emission from the 
halo position, which is then redshifted along the sightline.
In velocity space absorption can appear to occur ``behind'' the halo
due to the non-negligible width of the 
absorption profile, and because of infalling matter around the halo.

\subsection{Ly\texorpdfstring{$\alpha$}{[alpha]} scattering in the host halo}
\label{sec:exclusion_test}

The importance of carefully modelling the neutral gas in and close to the host halo
was emphasised and explored in \citet{2017ApJ...839...44S}, hereafter referred to as SZM17.
In Figure \ref{fig:n_HI_profiles} we show spherically averaged density profiles
for three mass bins spanning the masses of the halo population in our 
simulations. The neutral hydrogen densities were calculated assuming a fixed 
value of $\Gamma_\mathrm{HI}$ and solving for photoionisation equilibrium using 
Eq.~(\ref{eq:xeq}),
including the self-shielding prescription discussed in section \ref{sec:SS}.
Note however that spherically averaging will smooth out the overdensities
surrounding the halo which are used to calculate the amount of self-shielding;
this means that these radial profiles somewhat under-represent the neutral
gas density compared to sightlines through our simulations which are not 
spherically averaged. We see more extended profiles in the more 
massive haloes, whereas in the less massive halos the profiles are more peaked around the central 
halo position. In the central panel we also compare to profiles presented for
haloes with mass $M_h = 10^{10.5}\:M_\odot$ by SZM17. 
We note that the total hydrogen density
profiles are similar for $r>20$ pkpc, however at smaller radii both our total and neutral 
hydrogen densities are lower than the model presented by SZM17. This is
likely due to the \texttt{QUICK$\_$LYALPHA} star formation prescription, which converts dense
gas into star particles (as described in section \ref{sec:LSIS}). This prescription
will therefore remove some of the very dense gas in the centres of haloes, as 
we see in Figure \ref{fig:n_HI_profiles}.

We also show the neutral hydrogen profiles for the widely 
used self-shielding prescription proposed by \citet{2013MNRAS.430.2427R} with the 
thin (step) curves. These are  closer to those presented by 
SZM17, especially for the lower photoionisation rate of 
$\Gamma_\mathrm{HI}=10^{-14} {\rm s}^{-1}$. We suggest that 
most of the difference between the profiles in our simulations and the modelling 
of SZM17 is due the presence of ionizing 
sources. In the simulations on which the 
prescription of \citet{2017arXiv170706993C} is based, there are
ionizing sources within the self-shielded regions which affect the local
photoionization rate and therefore the self-shielding threshold density. 
These are not accounted for in the \citet{2013MNRAS.430.2427R} prescription.
Note that while SZM17 do account for a central ionizing source in their  
calculation, they assume this source to be rather weak.  
As already mentioned some of the difference will also be due to the 
spherical averaging which is not accounted for in our self-shielding prescription.
Note further that in this work we also consider the role of the larger scale 
ionization structure, and the 
presence of an IGM volume-filling neutral fraction, which SZM17  neglect.
As discussed by SZM17, the attenuation near to the host halo is very 
sensitive  to the distribution of neutral hydrogen close to the Ly$\alpha$ emitters. 
We discuss this in more detail in section \ref{sec:sadoun}.

CPBH15 and \citet{2013MNRAS.429.1695B}  did not attempt to simulate the
complex radiative transfer within the host halo, but instead assumed an 
intrinsic emission profile (for photons leaving the host system, but before 
attenuation by the IGM) and argued that this accounts for these effects. 
In those works the contributions of neutral gas within 20 pkpc were therefore 
neglected around the halo; for the narrower range of halo masses considered in 
those works this was a consistent and sufficient
exclusion. Our modelling here includes  a considerably larger range of 
halo masses, which therefore also have a considerable range of virial radii. 
Excluding gas within a fixed distance of 20 pkpc uniformly across our halo population
would remove all the 
neutral gas within a few virial radii around the less massive haloes, whilst 
only remove the gas within a fraction of the virial radius in the most massive 
haloes. Here  we therefore choose the exclusion region based on the mass
(or virial radius) of the host halo and will use our simulation and the 
$e^{-\tau}$ approach to account for the attenuation due to the neutral 
hydrogen in the outer part of the host haloes of Ly$\alpha$ emitters.  

We have tested the effect of varying the size of the exclusion region 
by excluding gas within 0.5, 1.0, 2.0, 5.0,
10.0 $R_\mathrm{vir}$, where we use $R_\mathrm{vir} = R_\mathrm{200,crit}$. 
The resulting transmission curves for these
exclusions, calculated as described in section \ref{sec:def_optdepth}, 
are shown in Figure 
\ref{fig:sadoun_transmission}. In the left-hand panels in shades of blue we
present the results for a sample of less massive haloes, whilst in the right-hand
panels in shades of red we show the results for more massive haloes. 
The important role of the gas peculiar velocities can be seen in the top panels
by comparing the solid lines (full calculations) to the dashed lines (calculated
neglecting peculiar velocities). In particular in the more massive haloes,
the peculiar velocities are sufficient to dramatically move the position
of the damping wing. We also note, considering the solid lines, that the
more massive haloes are more sensitive to the choice of exclusion: in the less
massive haloes (blue lines) the damping wing of the profile is moved by 
$\sim150$ km/s between the two exclusion extremes shown, whilst in the more 
massive haloes it is moved by $\sim350$ km/s. 

In the bottom panels of Figure \ref{fig:sadoun_transmission} we show the effect of
varying the chosen background  photoionisation rate $\Gamma$. This leads to a change in the amount
of equilibrium neutral hydrogen (self-shielded or recombined) within ionized
regions close to the halo. We see that for the higher photoionization rate
the effect of changing the exclusion region is reduced, and vice versa for the
lower photoionization rate.

Our fiducial choice is to exclude gas within 1.0 $R_\mathrm{vir}$; unless
otherwise specified, all results presented hereafter were calculated with this choice.
As can be seen in Figure \ref{fig:sadoun_transmission}, there will be some
dependence of the Ly$\alpha$ transmission on the chosen exclusion region. 
We mitigate this dependence with our choice of source models, as detailed
in section \ref{sec:Models}.
Further consequences of our choice of the size of the exclusion region 
are discussed in section \ref{sec:sadoun}.

\subsection{Transmission  fraction ratios (TFRs)}
\label{sec:define_TFR}

As we are primarily interested in the evolution of the Ly$\alpha$ attenuation 
during the epoch of reionization  we consider the ratio of 
transmission  fractions $T(z)/T(z_\mathrm{ref})$ (hereafter
referred to as TFRs), where $z_\mathrm{ref}$
is a reference redshift. In particular we choose to construct the ratio of 
higher redshifts with respect to $z_\mathrm{ref} = 5.756$, matching the choice
of $z=5.7$ common in the literature.

Narrowband (Ly$\alpha$-selected) observations of LAEs at different redshifts 
can be used to calculate the TFR evolution as \citep[][]{2017arXiv170501222K},
\begin{equation}
\label{eq:observed_TFR}
\frac{\widebar{T}(z)}{\widebar{T}(z_\mathrm{ref})} = 
\frac{\kappa(z_\mathrm{ref})}{\kappa(z)} 
\frac{f_\mathrm{esc,Ly\alpha}(z_\mathrm{ref})}{f_\mathrm{esc,Ly\alpha}(z)}
\frac{\rho_\mathrm{Ly\alpha}(z)/\rho_\mathrm{Ly\alpha}(z_\mathrm{ref})}
{\rho_\mathrm{UV}(z)/\rho_\mathrm{UV}(z_\mathrm{ref})},
\end{equation}
where $L_\mathrm{Ly\alpha} = \kappa L_\mathrm{UV}$, 
$f_\mathrm{esc,Ly\alpha}$ is the escape fraction of Ly$\alpha$ photons, and
$\rho_\mathrm{UV}$ is the intrinsic UV luminosity density whilst 
$\rho_\mathrm{Ly\alpha}$ is the observed (attenuated) Ly$\alpha$ luminosity
density. This relative transmission
fraction is an effective way of quantifying the evolution observed in the
LAE luminosity function. In particular it is a convenient quantity that
allows one to estimate the neutral fraction $x_\mathrm{HI}$ from an
observational sample. In this work we also choose to calculate the TFR
evolution rather than the luminosity function evolution because it can be
calculated via Eq.~(\ref{eq:transmissivity}) independently of the uncertain relationship
between the LAE host halo's mass and its Ly$\alpha$ luminosity. We leave the 
explicit modelling of the $M_h$-$L_\mathrm{Ly\alpha}$ relation, and hence the 
luminosity function evolution, to future work.

\subsection{Ly\texorpdfstring{$\alpha$}{[alpha]} Fractions}
\label{sec:EW}
Alongside the evolution of the Ly$\alpha$ luminosity function, observers have
also measured the evolution of the fraction of continuum-selected galaxies which emit strongly in
Ly$\alpha$. This is determined using samples of UV-selected galaxies (via the Lyman
break technique), with follow-up spectroscopy to measure Ly$\alpha$. The fraction, 
$X_\mathrm{Ly\alpha}$, is the proportion of such an LBG sample that are 
measured to have a Ly$\alpha$ equivalent width above a given threshold
\citep{2011ApJ...728L...2S,2012ApJ...744...83O,2012ApJ...747...27T}

In this work we also calculate the predicted evolution of $X_\mathrm{Ly\alpha}$
following a similar strategy to \citet{2017ApJ...839...44S}
and CPBH15. We start with the presription of \citet{2011MNRAS.414.2139D} in 
which we derive the \emph{rest-frame equivalent width}, $\mathcal{W}$, 
distribution. 
This is done by assuming that there is a probability distribution
$P_\mathrm{int}(>\mathcal{W})$ for an intrinsic unabsorbed $\mathcal{W}$ 
distribution which does not evolve 
with redshift; the observed redshift evolution is then entirely due to the 
attenuation by the IGM\@. Given the probability distribution for the transmitted 
fractions at a given redshift $P_{T}(T,z)$ and the instrinsic distribution, we 
can find the REW distribution at that redshift as,
\begin{equation}
P(>\mathcal{W},z) = \int_{0}^{1} \mathrm{d}T \: P_{T}(T,z) \: 
P_\mathrm{int}(>\mathcal{W}/T).
\label{eq:rew}
\end{equation}
As in CPBH15 we choose to determine $P_\mathrm{int}(>\mathcal{W})$ as the 
function which gives $P(>\mathcal{W},z=6)$ that matches the observational 
data of \citet{2011ApJ...728L...2S}. We fit the following functional form 
for the intrinsic distribution \citep{2003ApJ...588...65S},
\begin{equation}
    P_\mathrm{int}(>\mathcal{W}) = \exp(-\mathcal{W}/\mathcal{W}_0)/
    (\mathcal{W}_0 + \mathcal{W}_1),
\end{equation}
where $\mathcal{W}_0$ and $\mathcal{W}_1$ are free parameters which vary
depending on the simulated transmission fraction distribution. 
Given this intrinsic distribution,
and using Eq.~(\ref{eq:rew}), we can find the fraction of Ly$\alpha$ emitting
galaxies over a given threshold equivalent width as,
\begin{equation}
    X_\mathrm{Ly\alpha}(\mathcal{W},z) = P(>\mathcal{W},z).
\end{equation}
The values predicted by the simulations can then be compared to observed
fractions.


\section{Models}
\label{sec:Models}
Using the above simulation setup and Ly$\alpha$ transmission framework, we
can explore different models of reionization and LAEs to compare with
current observations. In particular we test three reionization histories
which bracket the possible progress of reionization at a given redshift. We also employ 
three different models for the masses of the host haloes of LAEs to explore 
the effect of host halo mass on Ly$\alpha$ transmission.
We therefore test a total of nine possible model combinations.
Further variations we have  considered are described in Appendix 
\ref{sec:C}.

\begin{figure*}
   	\includegraphics[width=2\columnwidth]{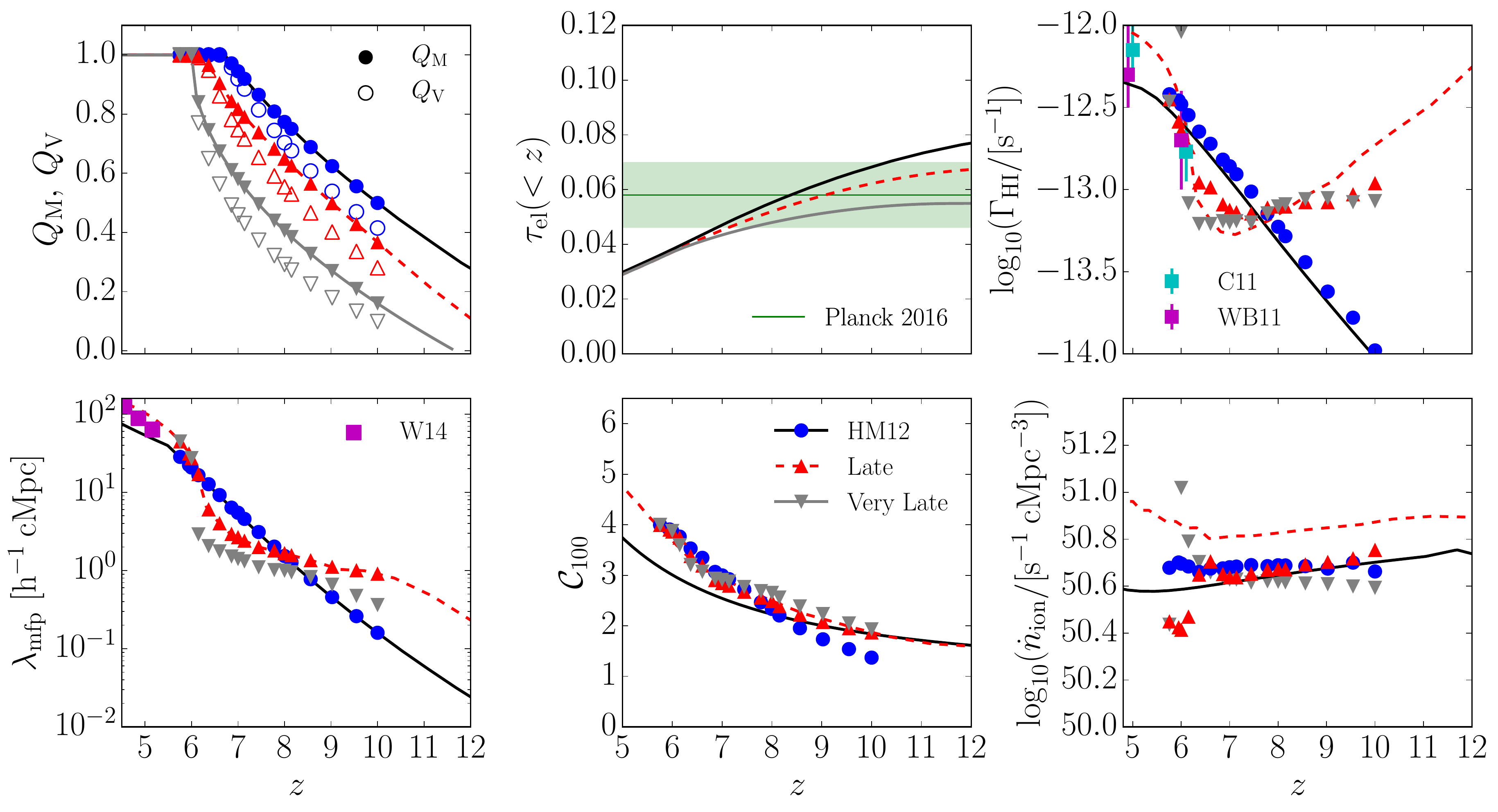}
    \caption{Calibrated parameters of the simulation.
        Clockwise from the top left: mass- and       volume-averaged 
        neutral fractions, electron-scattering optical depth,
        UV background photoionization rate, ionizing emissivity, clumping 
        factor for overdensities less than 100, ionizing photon mean free path 
        within ionized regions. 
        Our chosen models are shown as blue dots (HM12), red triangles 
        (Late) and grey inverted triangles (Very Late). 
        The reionization histories are shown as solid black 
        \citep[HM12]{2012ApJ...746..125H} and red dashed 
        \citep[Ch15]{2015MNRAS.453.2943C} lines. Observed data from 
        \citet[C11]{2011MNRAS.412.2543C}, \citet[WB11]{2011MNRAS.412.1926W}, 
        \citet[W14]{2014MNRAS.445.1745W} 
        are overplotted for comparison. Note that in the top middle panel 
        (electron-scattering optical depth) we show the
        1 $\sigma$ bounds shaded in green from \citet{2016A&A...596A.108P} 
    }
    \label{fig:params}    
\end{figure*}

\subsection{Reionization Histories}
We consider here three different reionization histories first discussed in CPBH15; we follow the naming 
convention established in \citet{2016MNRAS.463.2583K}. As outlined in section
    \ref{sec:calib}, each model provides $Q_M(z)$ and $\lambda_\mathrm{mfp}(z)$ which we input
into the calibration calculation.
\begin{itemize}
    \item \textbf{HM12}: this ionization history corresponds to the 
        commonly used 
        model of \citet{2012ApJ...746..125H}, based on the meta-galactic UV 
        background. We use $Q_M(z)$ and $\lambda_\mathrm{mfp}(z)$ as predicted
        in \citet{2012ApJ...746..125H}.
        In this model the galactic UV emission is used as a tracer
        of the cosmic star formation history; this can be derived from the
    galaxy UV luminosity function \citep{2013ApJ...768...71R}. Importantly
        the main contribution to the ionizing photon budget comes from 
        galaxies, with quasars and early Population III stars playing a
        negligible role. The universe is completely ionized in this model
        by $z=6.7$. Comparing the model predictions to observed data, it 
        agrees reasonable well with observed background photoionization 
        rates \citep{2009ApJ...703.1416F,2011MNRAS.412.2543C,
        2011MNRAS.412.1926W}. However its prediction for the Thomson optical 
        depth of the CMB, $\tau_\mathrm{el}=0.084$, is higher than the 
        measurement of \citet{2016A&A...596A.108P} by more than 1 $\sigma$.
    \item \textbf{Late}: this model uses the same evolution as the HM12 model
        with $Q(z)$
        shifted in $z$ such that reionization completes at $z=6$ instead of 
        $z=6.7$, but with the same $dQ_\mathrm{M}/dz$. 
        A similar reionization history was found in the full radiative transfer
        simulations of  \citet{2015MNRAS.453.2943C}, hereafter referred to as 
        Ch15. In Ch15 the radiative transfer code \textsc{Aton} 
        \citep{2008MNRAS.387..295A} was used to post-process  high resolution 
        cosmological hydrodynamical simulations calibrated to Ly$\alpha$ forest 
        data in order to calculate the evolution of the ionizing photon mean 
        free path. We use the mean free path predicted in that work for our
        calibration. The CMB Thomson optical depth is $\tau_\mathrm{el}=0.068$ 
        in this model.
    \item \textbf{Very Late}: Reionization completes
        at $z=6$ as in the Late model, but the evolution of $Q_\mathrm{M}$ is 
        much more rapid for $z>6$. AGN dominated reionization could
        lead to the history that this model predicts, see 
        \citet{2017arXiv170104408K} for further details. We predict the mean 
        free path for this model using the relationship
        between $Q_\mathrm{M}$ and $\lambda_\mathrm{mfp}$ from the Late 
        model\footnote{Beyond $z=6$, both of these quantities are
        monotonically increasing with redshift, and hence can be mapped 
        together. This allows us to find the mean free path for a given 
        $Q_\mathrm{M}$ of the Very Late model.}. The Thomson optical depth
        in this case is $\tau_\mathrm{el}=0.055$.
\end{itemize}
We follow \citet{2016MNRAS.463.2583K} in choosing the Late reionization 
history as our fiducial model.

In Figure \ref{fig:params} we show the final calibrated parameters of the 
simulation, including the reionization histories for $Q_\mathrm{M}(z)$. The
HM12, Late and Very Late calibrated parameters are shown as blue circles,
red triangles and grey inverted triangles, respectively, in all panels. The solid
black lines in all panels show the predictions of the underlying model from
HM12 \citep{2012ApJ...746..125H}, whilst the red dashed lines show the predictions
from Ch15 \citep{2015MNRAS.453.2943C}.
The fixed quantities are $Q_\mathrm{M}(z)$ and $\lambda_\mathrm{mfp}(z)$, shown on 
the left-most panels. We see reionization progresses from high redshift (where
$Q_\mathrm{M} \rightarrow 0$) until around $z \sim 6$; specifically in the HM12
model we see $Q_\mathrm{M}=1$ at $z=6.7$, whilst in the other models it reaches
1 at $z=6$. The optical depth of the CMB to electron scattering is 
predicted by the reionization history models, shown in the top middle panel. Here
the three lines for each model can be compared to the \citet{2016A&A...596A.108P}
value shown as a horizontal green line, with green shading indicating the $1 \: \sigma$ bounds. 
The quantities derived during our self-consistent calibration are the clumping 
factor (bottom middle panel), the ionizing emissivity (bottom right panel) and 
the background photoionization rate (top right panel). We see in the HM12 model
that the mean free path and the photoionization rate increase at a largely 
constant exponential rate as 
reionization progresses, with a roughly constant ionizing emissivity.
This smooth evolution of the mean free path may however be unrealistic 
\citep{2018arXiv180104931P}.
In comparison, the Late and Very Late models predict
a more steady photoionization rate at high redshifts, which suddenly increases
close to percolation at $z\sim7$ when the HII regions overlap to an extent 
that the mean free path of ionizing photons rises rapidly. For this more abrupt end to reionization to 
occur there needs to be a sharper increase in the mean free path, which can be seen in the
bottom left panel. We note that the recent physically-motivated model of 
\citet{2018arXiv180104931P} has been able to reproduce this required rather sharp 
increase.

\subsection{Host halo masses}
\label{sec:sources}

To model the effect of the IGM and CGM on the Ly$\alpha$ emission, 
we have to simulate  the underlying signal from the galaxies. This step of our simulation has two 
components: (i) the spatial distribution of galaxies in our simulation volume;
(ii) the emission profile, $J(\nu)$, of the galaxies. We expect the galaxy 
spatial distribution to follow the halo distribution \citep{1984ApJ...284L...9K,
2002MNRAS.335..432V}. Unfortunately the emission profile for high redshift galaxies 
is poorly constrained. Our modelling choices are motivated by the
tests discussed in section \ref{sec:exclusion_test}.

We consider three models for the spatial distribution of LAEs, based on 
different halo mass bins, choosing a sample of 4992 haloes per model. 
These models therefore have varying levels of 
correlation between the LAE positions and the positions/size of ionized 
regions.
\begin{itemize}
    \item \textbf{Small mass}: firstly we place the LAEs in haloes smaller than 
        the mean mass,
        which on average have a mass $M_h \sim 10^{9} \:\mathrm{M_\odot/h}$. 
        This simple model is useful for understanding the evolution of faint
        LAEs.
    \item \textbf{Large mass}: secondly we consider the case where LAE 
        positions have maximal correlation with the ionized regions, by placing
        them in the most 
        massive haloes of the simulation volume. These haloes have masses in the 
        range $10^{11} \lesssim M_h \lesssim 10^{12} \:\mathrm{M_\odot/h}$. 
        This model is used to represent the bright end of the LAE 
        distribution.
    \item \textbf{Continuous}: finally we place LAEs in a random sample taken 
        from the full halo population of the
        simulation, noting that the mass resolution of the simulation naturally
        enforces a physically realistic cutoff mass 
        $M_h > 10^{7} \mathrm{M_\odot}$ \citep{2017MNRAS.464.1633F}. Due to
        the steep slope of the halo mass function, this model will be dominated
        by smaller more common haloes, and hence will be similar in many respects
        to the small mass model. 
        It is intended as middle ground between the first two models, 
        and we consider it the most realistic model for comparing with an 
        observational survey of average LAEs.\footnote{Although we do not need
        to explicitely specify a 
        mass-luminosity mapping for the results in this work, 
        we note that  for the commonly assumed linear relation of Ly$\alpha$ luminosity 
        and host halo mass, $L_\mathrm{Ly\alpha} \sim {M_h}$, 
        the continuous model would correspond to a random sampling of the 
        faint end of the luminosity function.}
\end{itemize}
The first two models are used as approximate representations of the
different populations of faint (lower mass host haloes) and bright 
(higher mass host haloes) LAEs. 

\begin{figure}
   	\includegraphics[width=1.0\columnwidth]{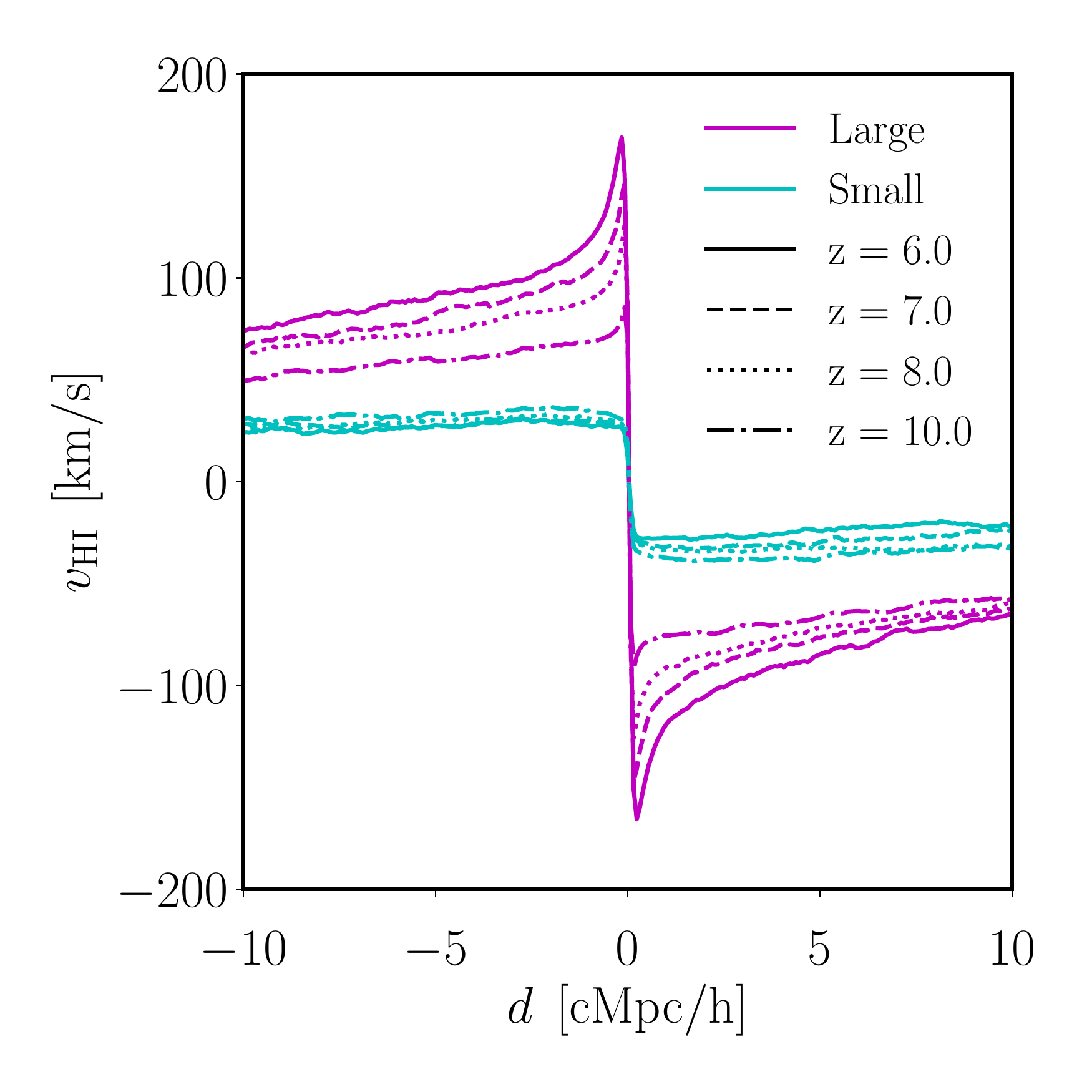}
    \caption{
        Median values of the hydrogen peculiar velocity 
        around haloes (at $d$ = 0 [cMpc/h]) are shown for the small mass
        range (cyan) and large mass range (magenta) for 5000 sightlines at
        $z=6$ (solid), $z=7$ (dashed), $z=8$ (dotted) and $z=10$ (dash-dotted).
        See Appendix \ref{sec:A} for a comparison of these simulation profiles
        with an analytical (excursion set) model.
    }
    \label{fig:sightlines}
\end{figure}

In Figure \ref{fig:sightlines} we show the 
median velocity for the gas distribution along sightlines through the
small mass (cyan lines) and large mass (magenta lines) haloes.
The figure shows much larger infalling velocities around the large mass 
haloes. Comparing across the different redshifts (with $z=10$
represented by the dash dotted lines, up to $z=6$ represented by the solid lines)
we also see more significant evolution in the larger mass 
haloes than for the smaller mass haloes. This evolution is largely driven by the
evolution in the halo masses of our large mass model, which can be seen in
Table \ref{tab:models}. Therefore our large mass model represents an upper limit
on the possible contribution the local gas environment evolution can provide towards 
Ly$\alpha$ attenuation.

We note that the peculiar velocities tend
to zero with increasing radius, but only on large scales of order ~80 cMpc/h. 
As a result of the long-range correlations of peculiar velocities, out to large radii from 
the host halo the gas is infalling with respect to the halo.
Comparing to the neutral gas density profiles in Figure \ref{fig:n_HI_profiles}
we see that the high column density gas around the more massive haloes will
be moved towards line centre (in the gas rest frame) by the large infalling
velocities.  Comparing these velocity profiles to the transmission curves in Figure
\ref{fig:sadoun_transmission} suggests  that there can be  increased  attenuation 
due to  damping wing absorption by  the neutral (self-shielded) gas around massive  
haloes compared to the less massive haloes of the small mass model.

\begin{table}
    \centering
    \caption{Averages masses of the different halo mass models used in this work.}
	\label{tab:models}
    \begin{tabular}{>{\tt}l|c|c|c|c}
		\hline \rule{0pt}{1em}
        \textnormal{Name} & \multicolumn{4}{c}{$\log_{10}(\widebar{M}_\mathrm{h}\:\mathrm{[M_\odot]})$}  \\
        \hline
               &      $z=6$ & $z=7$ & $z=8$ & $z=10$\\
        \hline
        Small mass  & $9.393$ & $9.358$ & $9.328$ & $9.283$ \\
        \hline
        Large mass & $11.531$ & $11.259$ & $11.002$ & $10.518$  \\
        \hline
        Continuous  & $9.594$ & $9.512$ & $9.477$ & $9.370$ \\
        \hline
	\end{tabular}
\end{table}

\begin{table}
    \caption{Estimated average host halo masses at $z=6.6$, using clustering 
    statistics like the angular correlation function (ACF).}
    \begin{tabular}{l|c}
        \hline \rule{0pt}{1em}
        Work &  $\log_{10}(\widebar{M}_\mathrm{h}\:/\mathrm{[M_\odot]})$ \\
		\hline
		\citet{2010ApJ...723..869O} & 10--11 \\
        \citet{2015MNRAS.453.1843S} & $\lesssim 10$ \\
		\citet{2017arXiv170407455O} & $10.8^{+0.3}_{-0.5}$ \\
		\hline
	\end{tabular}
    \label{tab:acf_masses}
\end{table}

\begin{figure*}
   	\includegraphics[width=2\columnwidth]{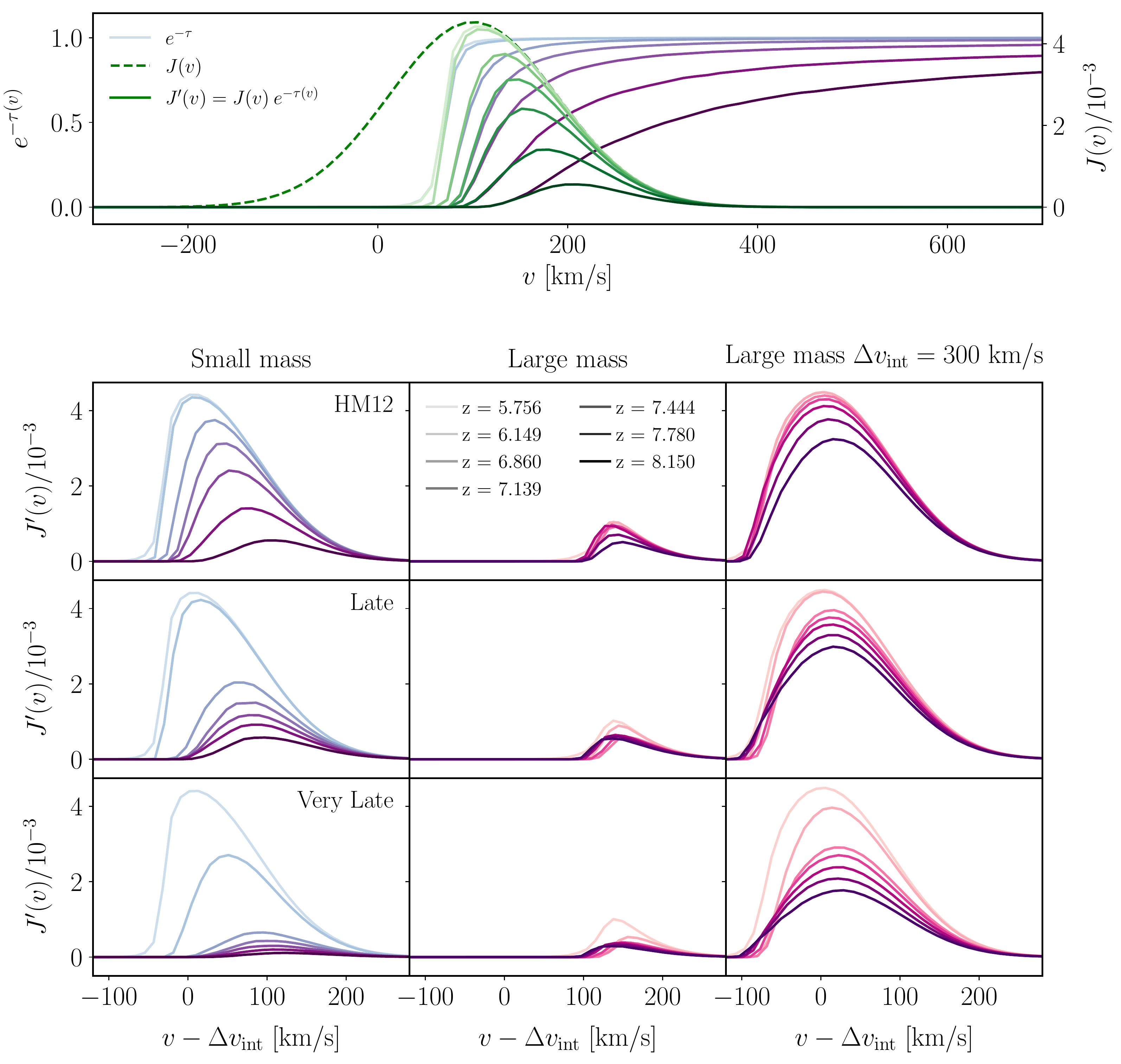}
    \caption{
        \emph{Top panel}: the (normalized) initial emission 
        profile $J(v)$ (assuming $\Delta v_\mathrm{int} = 100$ km/s, green 
        dashed line) and the median transmission 
        $e^{-\tau(v)}$ (purple solid lines) redward of line-centre 
        ($v = 0$), between $z=5.756$ (light) and $z=8.150$  
        (dark), are shown for the small mass model. The resulting 
        emission profiles (green solid lines) after IGM reprocessing are found as 
        the overlap of these two curves, $J' = e^{-\tau}J$. 
        \emph{Bottom panels}: the resulting median emission profiles for the 
        different mass and reionization history models. The small mass 
        host halo model is shown on the left and the large mass 
        model in the middle panels. The right panels also show the large mass
        results, but found using a larger intrinsic velocity offset of 
        $\Delta v_\mathrm{int} = 300$ km/s.
        The reionization histories (HM12, Late and Very Late) 
        are shown from top to bottom.
    }
    \label{fig:overlap}
\end{figure*}

For the second component of our source model we assume a Gaussian emission 
profile, with centre offset (in the galaxy rest-frame) from Ly$\alpha$ by a shift 
$\Delta \nu = \nu_\alpha \Delta v_\mathrm{int}/c$ , 
and width given by $\sigma_\nu = \nu_\alpha \sigma_v/c$. Importantly we account
for the peculiar velocity of the emitter when using the emission profile for
calculations in the frame of the sightline.
The radiative transfer through the ISM produces a characteristic double-peaked
emission profile \citep{2014PASA...31...40D}, however the blue peak will 
redshift into resonance while the photons traverse the IGM. At
the considered redshifts even residual neutral gas in ionized regions is 
sufficient to render this blue peak unobservable, 
hence our use of a singly-peaked Gaussian emission profile. It has been empirically
established that the Ly$\alpha$ emission line-centre is offset in both high 
redshift LAEs and lower-redshift analogs \citep{2015MNRAS.450.1846S,
2014ApJ...795...33E}. A suggested explanation for the cause of this offset is 
galactic outflows \citep{2010ApJ...717..289S,2014ApJ...788...74S},
but almost certainly in combination with resonant scattering effects 
\citep[e.g.][]{2011MNRAS.416.1723B}.  
We use the same values of $\Delta v_\mathrm{int}$ and $\sigma_v$ that were 
employed as the default model of CPBH15. The emission profile is the same for all
the haloes, with $(\Delta v_\mathrm{int}, 
\sigma_v) = (100, 88)\:\mathrm{km \: s^{-1}}$. These values are similar to those
inferred in \citet{2015MNRAS.450.1846S} using the \ion{C}{III]$\lambda1909$} line.

In summary we have nine model permutations, which include the three reionization
histories and the three halo mass models. Our fiducial model for comparison with
observational data is the `Late' reionization history combined with the continuous
mass model. In Appendix \ref{sec:C} we test
further model variations, including changes to the emission model (such as mass and 
redshift dependent velocity offsets).

\subsection{Observational constraints on host halo masses from LAE clustering}
\label{sec:hosts}
The best constraint   on host halo  masses of LAEs can be obtained using 
clustering statistics. The estimates for $z=6.6$ LAEs 
from \citet{2010ApJ...723..869O, 2015MNRAS.453.1843S, 2017arXiv170407455O}
are shown in Table \ref{tab:acf_masses}.
The average masses of host haloes have been calculated in the above works 
using samples that span the luminosity range from faint 
($10^{42} \lesssim L_\mathrm{Ly\alpha} < 10^{43}$ erg/s) to bright 
($L_\mathrm{Ly\alpha}\gtrsim 10^{43}$ erg/s), and so
do not necessarily reflect the expected masses for this distinction, but 
rather an average of the two ranges. We leave it to future work to perform a
detailed clustering analysis on the observed samples of LAEs split into these
luminosity brackets.
For comparison with this work, the average host halo masses at 
representative redshifts for our small and large mass models are shown in 
Table \ref{tab:models}.

Note again the definition of  our  mass models: large corresponds to 
the most massive haloes in the simulation, which evolves with redshift; 
small corresponds to the most common haloes with mass 
$\sim 10^{9}\:\mathrm{M_\odot}$. A comparison of Tables \ref{tab:acf_masses} 
and \ref{tab:models} shows that the observed masses lie somewhere in between our
small and large mass models. As mentioned in section \ref{sec:sources} our 
continuous model should thus be the  most representative of a real LAE sample.
Although the steepness of the halo mass function biases the average mass 
towards the smaller mass end of the spectrum, we still expect there to be LAEs 
hosted by the more massive haloes considered here.


\begin{figure}
   	\includegraphics[width=\columnwidth]{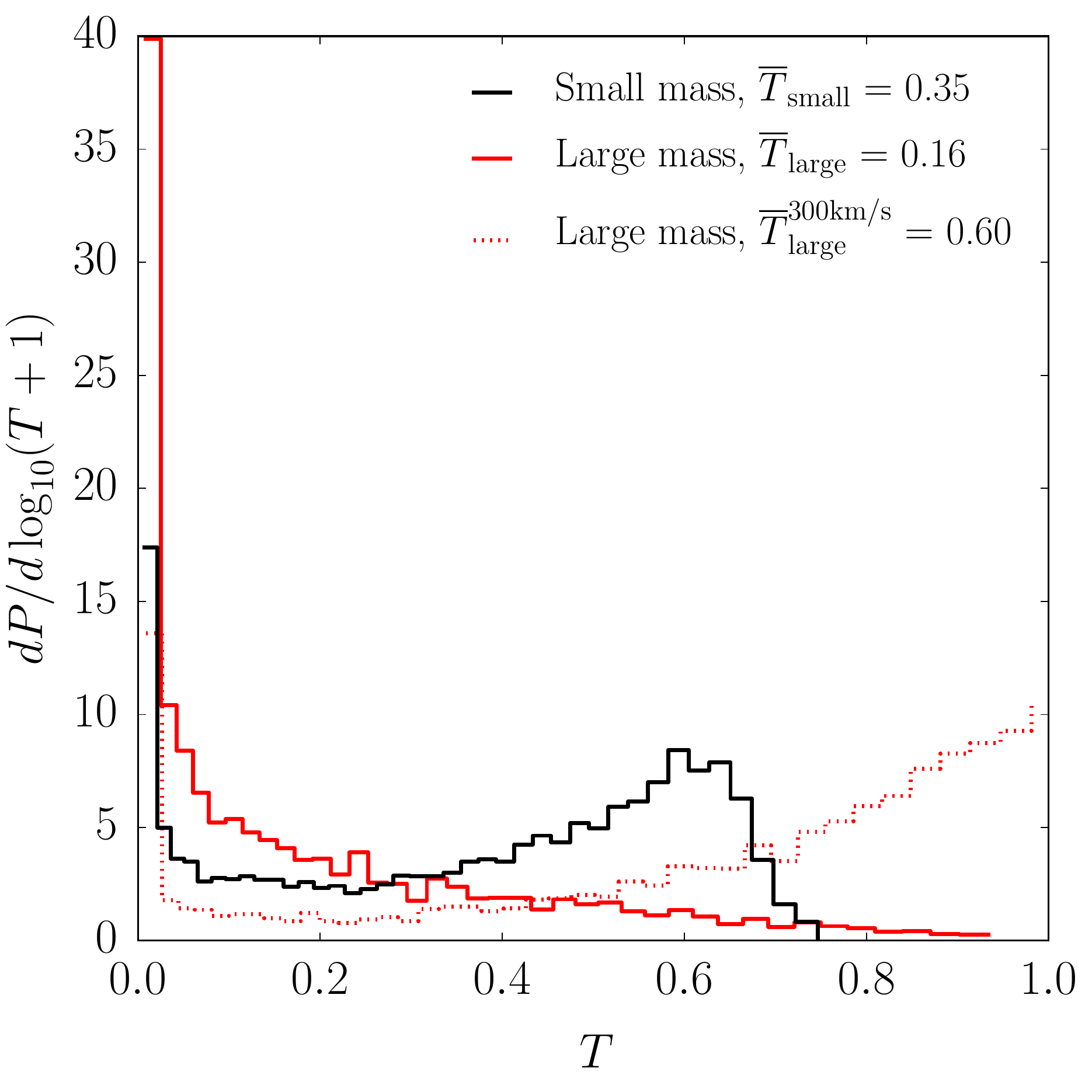}
    \caption{The distribution of transmission fractions for the small (black) 
    and large (red) mass models at $z=7$, using the fiducial 
    Late reionization history. The red dotted line shows the large mass model 
    distribution if the intrinsic offset is instead 
    $\Delta v_\mathrm{int}=300$ km/s.}
    \label{fig:T_dist}
\end{figure}

\begin{figure*}
   	\includegraphics[width=\textwidth]{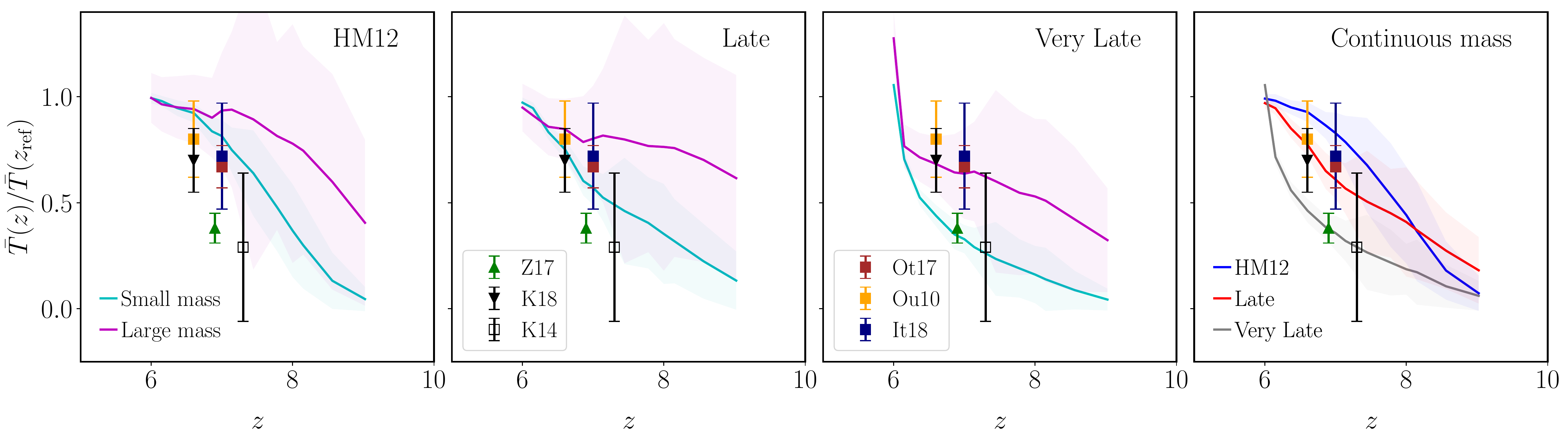}
    \caption{
        Evolution of the mean transmission fractions (TFRs), normalized to 
    $z_\mathrm{ref} = 5.756$.
    \emph{Lefthand three panels}:
     The large (magenta) and small (cyan) mass models
    are shown in all panels.
    The shaded regions show the 68\% scatter in the TFR values, found by sampling
    the distribution with sample sizes comparable to observational LAE samples at each redshift.
    From left to right we have the different reionization histories:
    HM12, Late and Very Late. Overplotted is data 
    from \citet[Z17]{2017arXiv170302985Z}, \citet[K18]{2017arXiv170501222K}, 
    \citet[K14]{2014ApJ...797...16K}, \citet[Ot17]{2017arXiv170302501O}, 
    \citet[It18]{2018arXiv180505944I} and \citet[Ou10]{2010ApJ...723..869O} 
    normalized to $z_\mathrm{ref}=5.7$; where
    errors were not quoted in these works we have made a basic estimate. Note 
    these observational data-points were found by considering the luminosity 
    density across the faint ($42 \lesssim \log_{10} L_\mathrm{Ly\alpha} 
    \lesssim 43$) and bright ($\log_{10} L_\mathrm{Ly\alpha} \gtrsim 43$) ends 
    of the luminosity functions. Therefore these observational points are best 
    compared to the continuous model, as shown in the rightmost panel.
    \emph{Rightmost panel}:
    The TFR evolution of the continuous mass model. The
    different reionization history models are shown in blue, red and grey, 
    with the corresponding shading indicating $1\sigma$ scatter. This 
    model represents a middle ground between the extreme small and
    large mass models. 
    Note that in all these panels the emission profile was our default model
    with $\Delta v_\mathrm{int}=100$ km/s.
}
    \label{fig:ratio_evol}
\end{figure*}

\section{Results}
\label{sec:results}
Having applied our calibration scheme for the different reionization 
history models, and then calculated the Ly$\alpha$ transmission for the 
different LAE models, we can now explore the effect of these different 
model parameters on the distribution of transmission fractions. We can also
explore the effect on the TFR (transmission fraction ratio, as defined in 
section \ref{sec:define_TFR}) evolution, and compare this to the observed
difference between bright and faint LAEs. Finally we can also derive the
evolution of the Ly$\alpha$ fraction, $X_\mathrm{Ly\alpha}$, and compare our
predictions with observations.

\subsection{Evolution of the median transmission}
In Figure \ref{fig:overlap}
we show the attenuation effect of the IGM on the initial galactic emission
profile. The top panel shows the components involved in the transmission fraction
calculation: the emission profile in dashed green, the transmission
in solid shades of purple (with shade darkening as redshift increases, for the small
mass model) and the
resulting transmitted emission profile (after IGM reprocessing) in solid shades
of green. 

The transmission  fraction is given by the area under this reprocessed
emission profile, as discussed in section \ref{sec:define_T}.
The lower set of panels show the reprocessed emission profile for 6 of the 
model combinations: the reionization histories from top to bottom, and the
small and large mass models in the left and middle panels respectively. 
We also show 3 further model
combinations in the right hand panels, in which the large mass model is paired 
with a larger intrinsic velocity offset of 300 km/s than our default 100 km/s.
In general the presence of neutral hydrogen gas during the EoR
causes the peak of the emission profile to be translated redwards in frequency space,
and to be reduced in amplitude. We note that the evolution of the profile
is most rapid in the Very Late model. For each reionization history it also
occurs more rapidly for the small mass model. The trend for the frequency 
translation of the profile with redshift is different between the small and 
large models. The small model profile reddens with increasing redshift. In the
large mass model the shift in frequency is less clear. We see that for the same
intrinsic emission profile, the resulting profile is more strongly attenuated
for the large mass  haloes at a given redshift. In the right panels
where we have used a larger intrinsic velocity offset 
($\Delta v_\mathrm{int} = 300$ km/s) we see that instead the large mass halo  
profiles are less (or equivalently)
attenuated compared to the small mass profiles. This demonstrates that the
IGM and CGM attenuation of the Ly$\alpha$ luminosity is indeed very sensitive to the 
intrinsic emission profile. Despite this significant effect seen when comparing
at a given redshift, we find that the relative transmission evolution (i.e. normalized
to a given reference redshift, as described in section \ref{sec:define_TFR}) 
is less  sensitive to the intrinsic emission profile. For further details see
Appendix \ref{sec:C}.

\subsection{Transmission fraction distribution}
\label{sec:TF}
We apply the framework from section \ref{sec:define_T} 
(Eq.~\ref{eq:transmissivity}) to explore the difference in the distributions of 
the transmission fractions for the small and large mass models. 
In Figure \ref{fig:T_dist} we show the (normalised) probability distribution 
for the transmission fraction at $z=7$; we show the small mass (black line)
and large mass (red line) models (as well as a large mass model with increased
velocity offset of 300 km/s shown with the dotted red line).
Considering first the small mass model distribution in black, we see a bimodal
distribution with peaks around $T\sim0$ and $T\sim0.6$. The $T\sim0.6$ peak 
can be understood as those sightlines which start in host 
haloes sitting in ionized regions, where there isn't sufficient recombined 
neutral hydrogen (or the neutral gas is not infalling with a high enough velocity)
to completely reduce the transmission fraction in the ionized region.
The photons emitted in the vicinity of such haloes can redshift beyond the 
damping wing by the time they reach the edge of the ionized region,
and hence will be transmitted along the sightline. The dominant $T\sim0$ peak is due to 
sightlines where photons emitted at the halo position would be
absorbed/scattered somewhere along the sightline. This absorption might be due
to self-shielded clumps, recombined hydrogen in the ionized regions, or 
residual neutral hydrogen in the rest of the IGM\@. Comparing this to the large
mass model distribution in solid red, we see instead a single peak around
$T\sim0$, although there is also a small non-zero probability of $T>0.8$ which wasn't
present in the small mass model distribution. Finally the red dotted line shows
the same large mass model, but using a larger intrinsic velocity offset of
$\Delta v_\mathrm{int}=300$ km/s (compared to the default of 100 km/s). This
distribution now recovers a second peak at $T\sim1$. We note that the mean 
transmission fraction is higher for the sightlines that start
on the small mass haloes, unless the larger velocity offset is used for the large
mass model haloes. 

These distributions may seem counterintuitive, as the more massive
haloes should sit in larger ionized regions and hence be more visible on average.
This picture however does not take into account the infalling velocities
of the neutral gas within ionized bubbles, either recombined or self-shielded,
which are considerably larger for the more massive haloes (as seen in Figure 
\ref{fig:sightlines}). This infall towards the halo counteracts the 
cosmological redshifting of the emitted photons such that they are closer to 
line centre in the frame of the gas, which leads to greater absorption 
(unless the intrinsic offset is increased). The $T\sim1$ peak in the default
large mass model (red solid line) is diminished because although these emitters 
sit in large ionized regions, the self-shielded gas within the ionized region 
can still strongly attenuate the Ly$\alpha$ emission. However when the intrinsic
offset is increased, such that this self-shielded gas becomes more transparent
to Ly$\alpha$ radiation, we recover the peak we would expect close to $T\sim1$.

In this way we see that at a given redshift the presence of neutral CGM gas
can lead to an increase in halo-to-halo scatter of the transmission in our mass samples. 
We note however that the average evolution of the transmission is driven by the
neutral IGM. The relative importance of the CGM/IGM absorption in Ly$\alpha$
visibility will be explored further in section \ref{sec:sadoun}.

\subsection{Transmission fraction evolution in the small, continuous and large mass models}
As discussed in section \ref{sec:define_TFR}, we can quantify the evolution
of the transmission  fraction by normalizing to a reference redshift value 
(here chosen to be $z=5.756$),
which we call the transmission fraction ratio (TFR). We calculate the mean TFRs
at a given redshift for the three mass models, in the three different
reionization histories. This can be used to compare how the visibility of LAEs
in the different mass models evolves. In Figure \ref{fig:ratio_evol} we plot
the TFR evolution of the small mass (cyan) and large mass (magenta) models, with
1 $\sigma$ scatter shown by the shading. We estimate this scatter by repeatedly 
sampling the transmission fraction distribution at each redshift, with sample sizes
comparable to the observational sample sizes\footnote{We note that for their luminosity function
samples, \citet{2017arXiv170501222K} found 1081 LAEs at $z=5.7$ and 189 at $z=6.6$.}. 
This results in an increase in 
scatter with redshift as the sample sizes decrease, reflecting the increase in
statistical uncertainty. Beyond redshift $z=7.3$ the
sample size is kept constant, and the scatter starts to decrease as the halo-to-halo
variation decreases (because at high redshifts the universe was more homogeneously 
neutral). In all reionization histories before 
percolation ($z \gtrsim 6$) we find that the large mass model evolves considerably slower than
the small mass model. We see greater scatter in the large mass model; this is
likely because the large mass model contains some rare very massive haloes,
such that there is a non-negligible difference in environment between the
most and least massive haloes within the large mass model. This
leads to more halo-to-halo variation in the Ly$\alpha$ transmission
along these sightlines, compared to the small mass model (whose mass bin width
is smaller).

Overplotted on Figure \ref{fig:ratio_evol} for reference are a 
selection of observed TFRs reported by \citet[Z17]{2017arXiv170302985Z}, 
\citet[K18]{2017arXiv170501222K}, \citet[K14]{2014ApJ...797...16K}, 
\citet[Ot17]{2017arXiv170302501O}, \citet[It18]{2018arXiv180505944I} and 
\citet[Ou10]{2010ApJ...723..869O}, all
normalized to $z=5.7$.
Importantly, the TFRs quoted by observers are usually calculated 
from a full (luminosity spanning) sample of LAEs, i.e.
\emph{including} both bright and faint LAEs. As the small and 
large mass models represent extreme examples of LAEs, the most meaningful 
comparison with observational data is with our 
continuous model. Nonetheless we overplot the observational data on
all TFR figures, in order to give a reference point for comparison.

In the rightmost panel of Figure \ref{fig:ratio_evol} we show the evolution of the TFRs for
the continuous model, with the different reionizaton histories represented by
different colours. We note that the ``Late'' (in red) and ``Very Late'' (in grey)
reionization histories are the best matches to the observed data, suggesting
a reionization history somewhere in between these two bracketing models.
The scatter in the continuous model is comparable
to the scatter in the small mass model, which in turn is similar to the
observational errors.

\subsection{Differential evolution of the transmission fraction}
\label{sec:diffTF}
Alongside the average TFRs reported by observers, some 
\citep[e.g.][]{2017arXiv170302985Z} have also reported that the TFRs for 
bright LAEs are higher than for faint LAEs. 
This behaviour is reproduced by our large (representing 
bright LAEs) and small (representing faint LAEs) models, 
which show a difference in the TFRs for the same redshifts as seen in
Figure \ref{fig:ratio_evol}.

\begin{figure}
   	\includegraphics[width=\columnwidth]{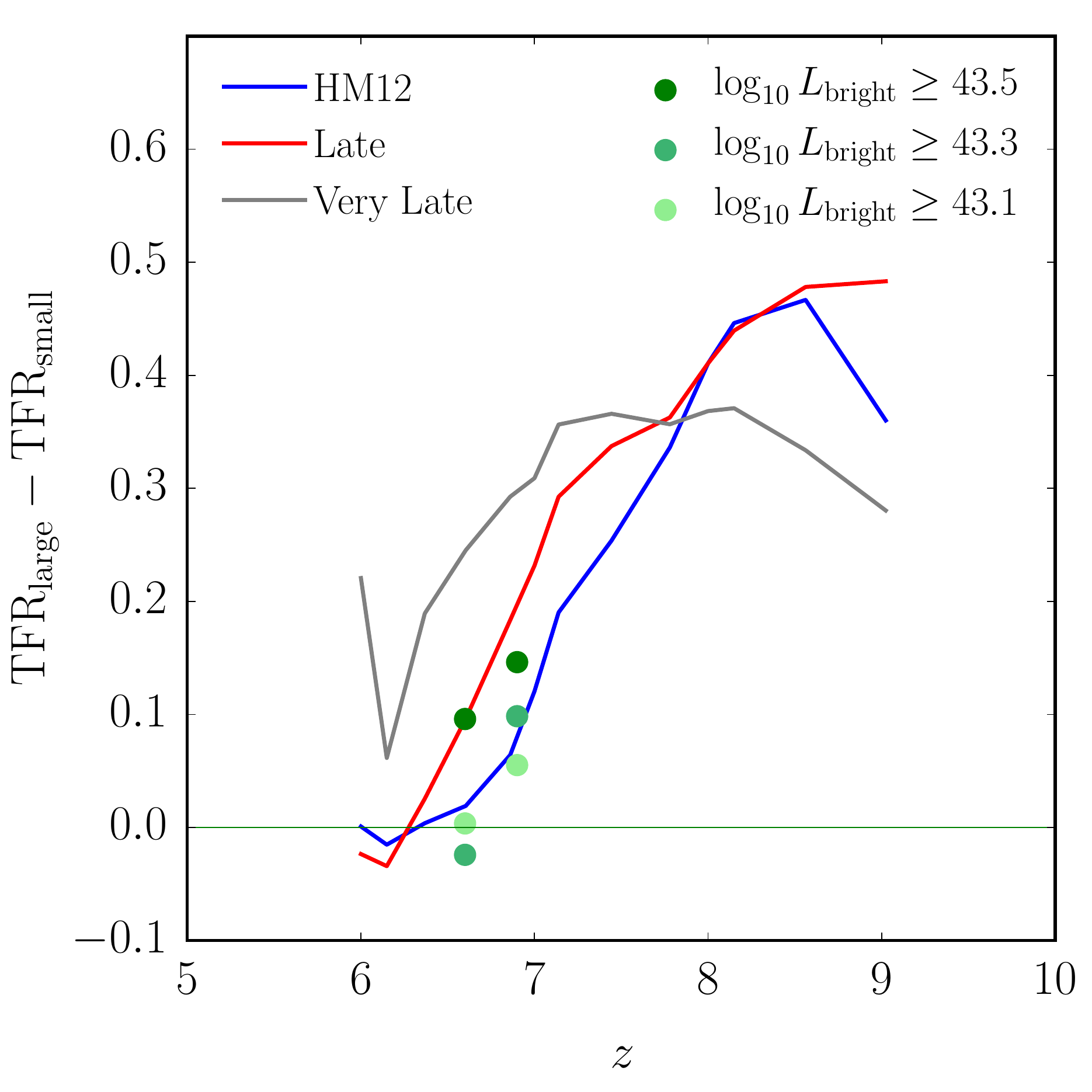}
    \caption{Difference in the evolution of the TFRs 
    (normalized to $z_\mathrm{ref} = 5.756$) between the large and small mass
    models. Overplotted in shades of green are observed differences, 
    derived with data
    from \citet{2014ApJ...797...16K, 2017arXiv170302985Z,2015MNRAS.451..400M,
    2016MNRAS.463.1678S} (this data was also normalized to $z_\mathrm{ref} = 
    5.7$) Three different brightness thresholds are shown: 
    $\log_{10}( L^\mathrm{bright}_{\mathrm{min}}/\mathrm{erg\:s^{-1}}) = 43.5,\:43.3,\:43.1$ .
    Note again that in all these models the emission profile was our default model
    with $\Delta v_\mathrm{int}=100$ km/s.
    }
    \label{fig:ratio_evol_data}
\end{figure}

\begin{figure*}
    \centering
   	\includegraphics[width=\linewidth]{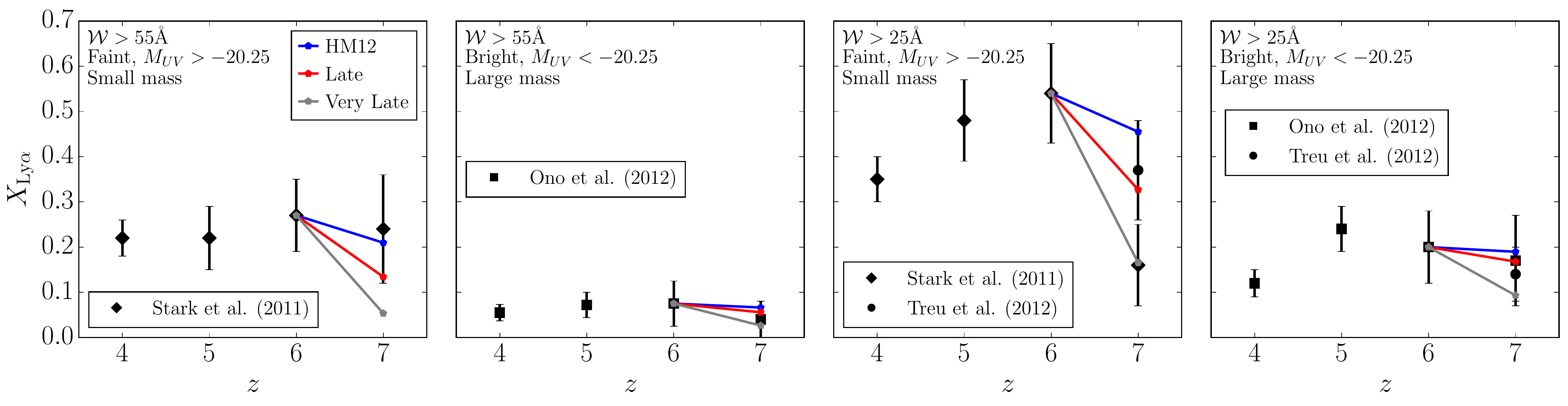}
    \caption{The fraction of galaxies emitting Ly$\alpha$ above a given
threshold equivalent width $\mathcal{W}$ (EW), as a function of redshift. 
These distributions were calculated using the small mass model (first and third panels, UV faint)
and large mass model (second and fourth panels, UV bright).
The different reionization models are shown as blue (HM12), 
red (Late) and gray (Very Late) lines. 
For comparison the observed \emph{UV faint} ($M_\mathrm{UV} > -20.25$) data 
from \citet[][diamonds]{2011ApJ...728L...2S}, \citet[][circles]{2012ApJ...747...27T} 
and \emph{UV bright} ($M_\mathrm{UV} < -20.25$) data from 
\citet[][squares]{2012ApJ...744...83O}, \citet[][circles]{2012ApJ...747...27T} are 
also shown. }
\label{fig:REW_dist}
\end{figure*}

To compare this more explicity, we use reported observational data
from \citet{2014ApJ...797...16K,2017arXiv170302985Z,2015MNRAS.451..400M,
2016MNRAS.463.1678S} to reconstruct the TFRs for the bright and faint LAEs
separately. In this way we want to establish trends and obtain a lower limit 
on the bracketing values 
for the TFRs, and so do not perform a detailed re-analysis of the data.
We take values of $\Phi(L) = {dn}/{d\log_{10}L}$ as quoted in the original works. 
From these we calculate,
\begin{equation}
\rho(z) = \int_{L_\mathrm{min}}^{L_\mathrm{max}} L \:\Phi(L) \:d\log_{10}L.
\end{equation}
In order to perform this integral we apply a trapezoidal algorithm on the 
published data points; we do not fit a Schechter form. 
The data is heterogeneous in terms of the luminosity ranges observed, so we
impose limits, $L_\mathrm{Ly\alpha} \in [42.5,43.7]$ erg/s, and use linear
interpolation and extrapolation to evaluate each of the datasets in the same
luminosity bins along this range.
There is obviously freedom in the choice of the ``bright'' threshold; 
we test values around $\log_{10} L^\mathrm{bright}_{\mathrm{min}} = 43$,
bracketing $\log_{10} L^\mathrm{bright}_{\mathrm{min}} = 43.4$ as used in 
\cite{2017arXiv170302985Z}. The threshold used for the calculated values plotted in 
Figure \ref{fig:ratio_evol_data} are 
$\log_{10}( L^\mathrm{bright}_{\mathrm{min}}/\mathrm{erg\:s^{-1}}) = 43.1,\: 43.3,\: 43.5$.
We then calculate the TFRs using the expression in Eq.~(\ref{eq:observed_TFR}),
with the UV data from \citet{2015ApJ...810...71F}\footnote{We also calculated 
using UV data of \citet{2015ApJ...803...34B}, but the bright/faint trend 
persists regardless of this change.}, assuming $\kappa(z=5.7)/\kappa(z=6.6)=1$
and $f_\mathrm{esc,Ly\alpha}(z=5.7)/f_\mathrm{esc,Ly\alpha}(6.6)=1$.

The green markers in Figure \ref{fig:ratio_evol_data} show the 
(re-calculated) difference in the TFRs for the bright and faint LAEs, 
compared to the simulated differences between the large and small mass models 
(lines coloured by reionization history).  The differential TFR evolution depends
on the chosen reionization history, but the shape of this evolution 
is similar across the models. We  will discuss this  further in section
\ref{sec:bubeffect}. Note that changing the bright threshold in the observed
data alters the amplitude of the difference, and the slope across redshifts.
\footnote{We note that the simulated TFR difference is sensitive to the chosen
mass bins, and hence without better constraints on host halo masses the
observed TFR differences cannot be used to constrain the
most likely reionization history.  Our large and small mass models 
are nevertheless useful for demonstrating  that a difference does indeed 
occur.}

\subsection{Evolution of the Ly\texorpdfstring{$\alpha$}{[alpha]} fraction of LBGs}
Finally, we also consider the independent observational measurement of the Ly$\alpha$
fraction of LBGs, $X_\mathrm{Ly\alpha}$, to see if our large and small mass models can be
used to reproduce the UV bright and UV faint evolution. We calculate this
evolution as described in section \ref{sec:EW}.

In Figure \ref{fig:REW_dist} we compare the evolution of $X_\mathrm{Ly\alpha}$
predicted by our simulations with the observed data, for the
thresholds of $\mathcal{W}>25$ \r{A} and $\mathcal{W}>55$ \r{A}. 
Our models are again reasonably consistent with the data; the largest discrepancy
is found for the steep drop in the $\mathcal{W}>25$ \r{A} UV faint data of 
\citet{2011ApJ...728L...2S} which only our `Very Late' model is able to reproduce. We note
that the use of the large mass model for the UV bright data accounts
for the slower decline in this sample, whilst the faster evolution of
our small mass model is a good fit for the UV faint sample. Apart from
the left panel ($\mathcal{W}>55$ \r{A} UV faint), the comparison with
observational data does not exclude any of our reionization history 
models.

Comparing with SZM17 \citep{2017ApJ...839...44S}, we see similar predictions
to their infall model, despite having a more modest evolution of 
$\Gamma_\mathrm{HI}$\footnote{Their model considers a change between $z=7 
\rightarrow 6$ in the photoionization rate of $\Gamma_\mathrm{HI} = 10^{-14}
\rightarrow 10^{-13}\:\mathrm{s}^{-1}$; the minimum $\Gamma_\mathrm{HI}$ in 
our models between $6<z<8$ does not fall below $10^{-13.2}\:
\mathrm{s}^{-1}$.} in our simulations.


\section{Discussion}
\label{sec:discussion}
In section \ref{sec:diffTF} we have shown  that our simulations 
predict a difference in the evolution of the visibility of LAEs hosted in
different mass haloes. 
If we assume that indeed brighter LAEs are found in more massive host haloes, 
then this  can explain the different  evolution of bright and faint 
high-redshift LAEs. 
We now discuss possible physical mechanisms for this difference in our 
simulations. We caution, however, that some of the observed difference could 
also be due to observational selection effects.

\subsection{Differential evolution of large and small mass models}
\label{sec:bubeffect}
Neglecting intrinsic galaxy evolution, we  explore here two different aspects of
the IGM and CGM attenuation that might cause the different evolution of the bright and 
faint LAE populations. 
\begin{enumerate}
    \item The most intuitive mechanism is perhaps the different (large-scale) 
enviroments  of ionized bubbles in which LAEs might reside 
\citep[for example suggested by][in section 4.1]{2017arXiv170302501O}. More massive haloes
are likely to reside in larger ionized regions compared to less massive haloes.
In particular we might also expect that (depending on the reionization 
history) more massive haloes will be surrounded by ionized regions earlier,
after which their visibility will not evolve dramatically; in comparison the
less massive haloes will enter  overlapping  ionized regions around the more 
massive haloes at later times.
    \item A second, more subtle mechanism is due to the different dynamical properties of 
neutral hydrogen in the CGM. In Figure \ref{fig:sightlines} we showed
the evolution of the infall velocity  of gas around haloes of different masses. 
We might expect both the gas close to the halo (which includes self-shielded or recombined 
neutral hydrogen within the ionized region) and the residual neutral gas in the
not-yet-ionized IGM around the halo to absorb differently depending on the host
halo mass.
\end{enumerate}

\begin{figure*} 
    \includegraphics[width=2\columnwidth]{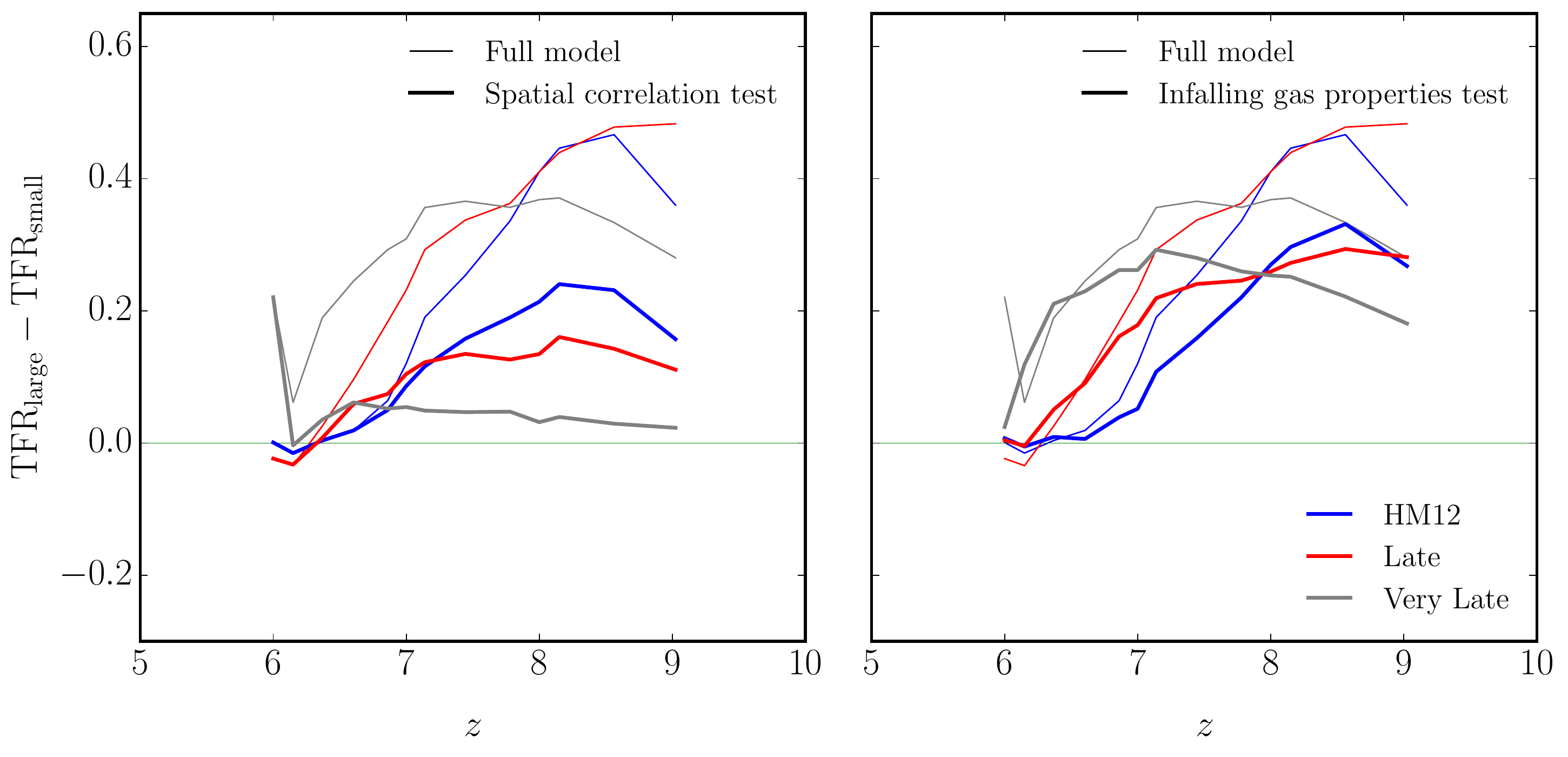}
    \caption{Evolution of the ratio of transmission fractions, testing the 
    effect of the CGM/IGM on the differential visibility of large and small host 
    haloes. The difference in TFR between the large and small mass models is 
    plotted for: the full models (thin lines), the test cases
    (thick lines). In the left panel we show the ``Spatial correlation test'', 
    whilst in the right panel we show the ``Infalling gas properties test''.
    The different reionization history models
    are shown in blue (HM12), red (Late) and gray (Very Late).
    }
    \label{fig:ratio_evol_test}
\end{figure*}

In order to explore the contributions of these two mechanisms in our models,
we perform the following two  tests.
\begin{itemize}
\item \emph{Spatial correlations with ionized bubbles}: we displace the 
    ionization field along one 
    direction of the simulation by a distance $d=80$ cMpc/h,
    half the simulation box length. This will break the correlation between 
    the location of haloes and ionized regions. 
    If the higher TFRs of the more massive host haloes are caused by their position in 
    larger ionized bubbles, 
    then this test should result in the TFRs of the small mass and large mass models 
    converging.
\item \emph{CGM peculiar velocities and temperature}: we recalculate the optical 
    depths along the extracted sightlines, but neglect both the
    peculiar velocities and the temperature variation of the CGM. 
\end{itemize}
Note, however, that the two mechanisms are coupled and that the two tests 
are therefore not independent of each other. In particular in the first test,
by displacing the ionization field we will also be removing some of the 
correlation between the in-falling velocities of neutral gas and the haloes.
We should therefore not expect the  two tests to quantify how much each of the 
mechanisms is contributing to the difference in TFR evolution, but they should nevertheless  
show whether these two mechanisms are indeed having an effect.

In Figure \ref{fig:ratio_evol_test} we plot the difference between the TFRs for 
the large and small mass models as a function of redshift. 
The left panel  shows the spatial correlation test.  
The difference drops close to zero for  all reionization histories suggesting 
that indeed the difference  of  the visibility of smaller and large mass haloes   
decreases significantly if there is no correlation of their location 
with that of ionized bubbles. We note that the effect of the correlation depends
on the reionization history.  The strongest effect of removing the correlation 
is seen in the Very Late model  and it is weakest for  the HM12 model. 

This can be understood by considering the rate with 
which smaller haloes enter the large-scale overlapping ionized regions. 
The overlapping ionized regions initially 
develop around the largest haloes which provide a bigger fraction of the total ionizing
photon budget, and hence these haloes remain in ionized regions out to higher redshifts. 
Smaller haloes enter into these ionized regions later, when the ionization fronts
around the larger haloes percolate and expand into the ionization fronts around
these smaller haloes. How quickly the smaller haloes enter the ionized regions
depends strongly on the reionization history, both on $Q_\mathrm{M}(z)$ and
$dQ_\mathrm{M}/dz(z)$. In the  HM12 model  reionization ends early
such that around $z\sim7$ both the small and large halo positions are strongly
correlated with the ionized regions. This means the difference in visibility
of these haloes is mostly not determined by the sizes of the ionized bubble. In 
the Very Late model, in which reionization ends later
and more rapidly, there is a much larger difference in the bubble sizes
surrounding the small and large haloes at  $z\sim7$. 

In the right panel Figure of \ref{fig:ratio_evol_test}  we show the effect of 
neglecting gas peculiar velocities 
and temperatures of the CGM surrounding the host haloes. 
Neglecting these gas properties, there will be less 
absorption in neutral hydrogen around the host halo (both within reionized regions
and also in the residually neutral IGM). For this test there 
is also a dependence on the reionization history,
however for all our reionization models the effect is less significant than that
of removing the spatial correlation with the ionized bubbles. 
The influence of the infalling gas properties increases with redshift. 
For example the fractional difference from the
full calculation in the Late model is $\sim0.3$ at $z=8$, but rises to $\sim0.4$ by
$z=9$.

In summary, the results of our two tests suggest that, dependent on the 
reionization history:
\begin{itemize}
    \item the positions of more massive haloes in larger ionized regions can 
        make a significant contribution to the differential visibility of the 
        large and small mass models.
    \item the infalling gas properties of neutral IGM gas also play an, albeit smaller,  
        role in the increased visibility of the large mass model.
\end{itemize}

We also note that the largest difference in visibility occurs for  the 
Late reionization history.

\subsection{The effect of self-shielding and the dominant scales on which IGM 
attenuation occurs}
\label{sec:sadoun}

In SZM17 \citep[and also in earlier work such as][]{2016MNRAS.463.4019K,2007MNRAS.377.1175D}, 
the role that the infalling CGM gas plays in the Ly$\alpha$
attenuation was explored. We have seen in Figure \ref{fig:sadoun_transmission}, 
that the self-shielded gas in the CGM can indeed attenuate 
Ly$\alpha$ alongside the more distant neutral gas in the not yet ionized regions 
of the IGM. The strength of
the attenuation depends on the amount of self-shielded gas present, and
hence also on the local photoionization rate. As the global neutral fraction of
the large-scale IGM is also coupled to the photoionization rate, we note that
these two attenuating components are also coupled.
Within our models, the strength of the attenuation due to the self-shielded
gas in the CGM will depend on the assumed self-shielding prescription,
the amount of gas that is excluded from within the host halo, and the intrinsic
velocity offset of the Ly$\alpha$ emission profile. In this subsection we aim
to explore the interplay between this inner CGM self-shielded gas and the 
external (residual) neutral IGM gas, to try to quantify the strength of the roles
that they play in attenuating Ly$\alpha$ from high redshift galaxies.

\begin{figure*}
    \centering
   	\includegraphics[width=\linewidth]{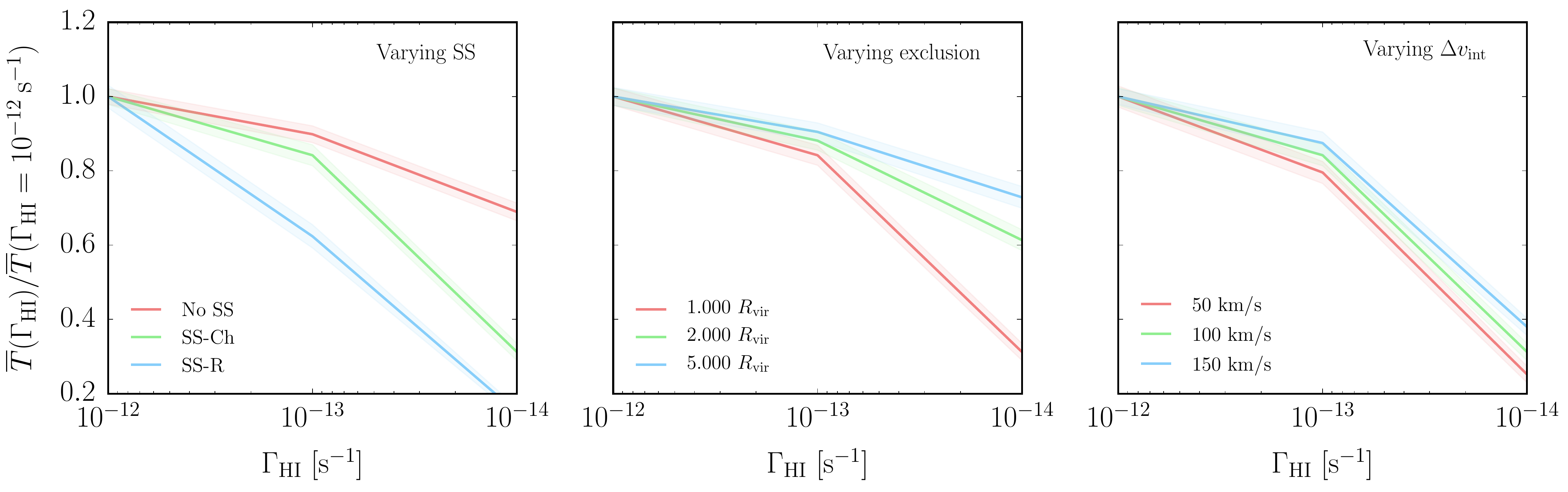}
    \caption{The photoionization rate dependence of the transmission fraction at 
    $z=6$, normalized to the value when $\Gamma_\mathrm{HI} = 10^{-12}$ s$^{-1}$.
    \emph{Left}: The effect of varying the self-shielding prescription.
    \emph{Middle}: The effect of varying the amount of gas that is excluded 
    around the halo position.
    \emph{Right}: The effect of varying the intrinsic velocity offset of the
    emission profile. Shading indicates 68\% scatter around the mean, calculated
    as in Figure \ref{fig:ratio_evol}.
    }
    \label{fig:tfr_g_ss}
\end{figure*}

In Figure \ref{fig:tfr_g_ss} we show how the transmission fraction at $z=6$
depends on the background photoionization rate. In all of our reionization
histories at this redshift, the IGM is ionized ($Q_\mathrm{M}=1$), and so
only the self-shielded/recombined CGM gas can play a role. In order to quantify
how strong the attenuation can be from this gas, we normalize the transmission
fraction to the value for $\Gamma_\mathrm{HI}=10^{-12}$ s$^{-1}$.
In each of the three panels of Figure \ref{fig:tfr_g_ss} we then test the effects
of our assumptions: on the left the self-shielding prescription, in the middle
the exclusion regime and on the right the emission profile offset.
In all panels we see that decreasing the background photoionization rate
(and therefore increasing the amount of self-shielded gas) increases the 
attenuation of Ly$\alpha$. We note that in our fiducial reionization history,
however, the background photoionization rate doesn't fall lower than 
$\Gamma_\mathrm{HI} \sim 10^{-13.2}$ s$^{-1}$.

In this work we have employed the self-shielding prescription suggested by
\citet{2017arXiv170706993C} (labelled SS-Ch). Other works have used different prescriptions
for self-shielding, which can lead to more neutral gas and thus a stronger
attenuation of Ly$\alpha$ emission. In the left panel of Figure \ref{fig:tfr_g_ss} we compare our 
fiducial prescription with: (i) the case of no self-shielding, and (ii) with a 
prescription based on \citet{2013MNRAS.430.2427R} (the default choice of CPBH15,
labelled SS-R).
As expected we see that the stronger the self-shielding the more attenuation
can come from this CGM gas. However even in the case of no self-shielding, 
where the amount of neutral gas is given only by recombinations in photoionization
equilibrium, we see that if the photoionization drops sufficiently then the
transmission fraction can be reduced.
In the middle panel we see the effect of excluding different amounts of the CGM
gas. Importantly we note that for our default SS-Ch self-shielding prescription and photoionization
rates larger than $\Gamma_\mathrm{HI} \sim 10^{-13}$ s$^{-1}$ (that are suggested
by full radiative transfer simulations of reionization), the attenuation
is not very sensitive to the size of the exclusion region.
In the righthand panel we show the effect of changing the intrinsic velocity 
offset of the emission profile. We see that of the three assumptions tested in this figure,
the results are least sensitive to this choice.

Note that, for the photoionization rates in our 
reionization histories, the self-shielded CGM gas alone can attenuate the 
Ly$\alpha$ signal by as much as $\sim30\%$ for the \citet{2013MNRAS.430.2427R} 
self-shielding. For this model the dependence on the size of the exclusion 
region is therefore also stronger than our default self-shielding model.

\begin{figure*}
   	\includegraphics[width=\linewidth]{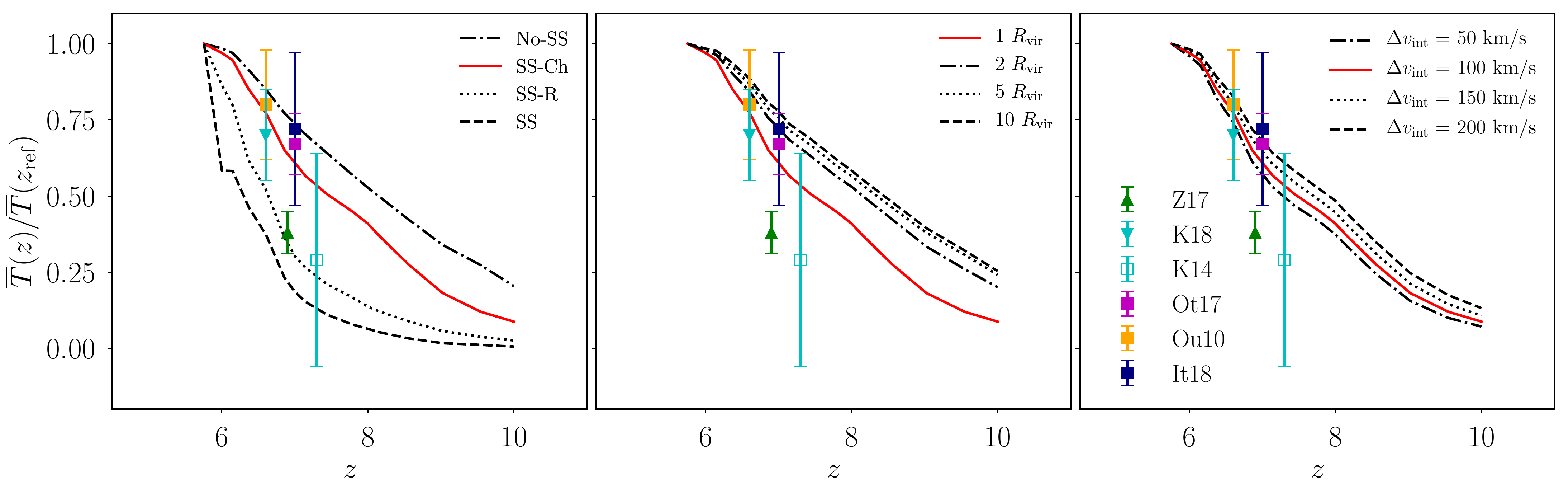}
    \caption{Testing the effects of different assumptions on the TFR evolution.
    \emph{Left}: The effect of a range of self-shielding prescriptions, 
    using the Late reionization history and the continuous
    model. The fiducial prescription used in this work is the
    SS-Ch model based on \citet{2017arXiv170706993C}, shown in red.
    \emph{Middle}: The effect of excluding gas around the halo, within a number
    of virial radii. Our fiducial exclusion is 1 $R_\mathrm{vir}$, shown in red.
    We note that beyond an exclusion of 2 virial radii, the evolution
    is very insensitive to the exact choice of exclusion radius, 
    however there is still a strong attenuation. This remaining 
    attenuation is due to the large-scale neutral IGM, as previously modelled.
    \emph{Right}: The effect of changing the intrinsic velocity offset of the 
    emission profile. Our fiducial offset is 100 km/s, shown in red.
    Overplotted are observed TFR values, as in Figure \ref{fig:ratio_evol}. }
    \label{fig:altSS}
\end{figure*}

In order to explore this further, we also show the effect of changing our
assumptions for the full TFR evolution in Figure \ref{fig:altSS}, using
the continuous mass model and the Late reionization history. This
therefore includes the contributions of both the CGM and the IGM. As in the 
previous figure, on the left panel we show the effect of the self-shielding
prescription, in the middle we show the effect of the exclusion region, and in
the right panel we show the effect of the velocity offset. In the left panel
we also include the prescription used in \citet{2013MNRAS.429.1695B} (labelled SS),
which assumes a sharp threshold for self-shielding at the Jeans scale. The
results found without self-shielding can be considered as the attenuation
due to the residual neutral IGM alone. We see that the neutral IGM is the 
dominant component in determining the average redshift dependence of the
attenuation. However the self-shielded gas can also play an important
role, depending on the self-shielding prescription (SS resulting in the most
self-shielding, and SS-Ch the least).
In the central panel we see the effect of excluding different amounts of the CGM
gas. For exclusion regions $> 2 R_\mathrm{vir}$, the TFR depends very weakly on 
the exact choice of exclusion radius and values close to those in the No-SS case
(shown in the left panel) are found.
Finally we see in the right hand panel that varying the intrinsic velocity
offset does not alter the TFR evolution very much. Although the transmission
at a given redshift might be sensitive to these changes, the normalization of the
TFR removes part of this sensitivity (so long as the velocity offset is 
independent of redshift).

\subsection{Observational selection effects}
\label{sec:selection}
Throughout this work we have relied on the basic assumption that there is a
positive correlation between the host halo mass and a galaxy's (rest-frame)
Ly$\alpha$ luminosity. 
Ly$\alpha$
photons are created in a galaxy's ISM by reprocessing the ionizing photons
emitted from the stellar component. The Ly$\alpha$ luminosity 
depends on the star formation rate (SFR), which in turn depends on the host 
halo mass, $M_h$, \citep{2010ApJ...716..574Z}. Given the often bursty nature
of star formation it is nonetheless not obvious that the brightest LAEs are 
hosted in the most massive haloes.

In the first instance we have calculated the TFR evolution, and 
compared to narrowband Ly$\alpha$ selected galaxies 
\citep[such as in][]{2010ApJ...723..869O}. 
We split the samples into \emph{bright} and \emph{faint} based on the observed
Ly$\alpha$ luminosity. For this selection method a galaxy might
be categorized as a bright LAE but might not necessarily be hosted by a more
massive host halo. This is because the flux in the Ly$\alpha$ narrowband filter
is compared with a (sometimes overlapping) broadband filter; the galaxy may
appear bright with this selection method because there is more flux in the
narrowband than in the UV continuum. This therefore includes cases where the
UV continuum is faint, and hence the galaxy may be less massive. 

We have also calculated the evolution of 
$X_\mathrm{Ly\alpha}$, and compared to dropout selected galaxies with 
spectroscopically confirmed Ly$\alpha$ equivalent widths ($\mathcal{W}$) above 
a given threshold \citep[such as in][]{2012ApJ...744...83O}. These galaxies are
first selected using the Lyman break technique, and divided into UV-bright and
UV-faint, based on bolometric UV luminosity. This UV luminosity correlates with
stellar mass, and hence the UV-brighter objects will be hosted in larger mass
haloes. The secondary Ly$\alpha$ equivalent width selection does not change this
measurement, so in this case the brighter LAEs will almost certainly correspond 
to more massive haloes.  
\citet{2012MNRAS.421.2568D} have suggested that indeed the $z>6$ 
LAEs form a luminous subset of LBGs.

The applicability of our different mass models, and in particular the mapping
from these models to the different populations of LAEs (divided by brightness),
is therefore dependent on the way the population is selected. The TFRs we have 
calculated using the continuous mass model are probably the most realistic. For the 
$X_\mathrm{Ly\alpha}$ evolution however, our application of the different mass
models to the different UV brightness samples is probably better justified.


\section{Conclusions}
\label{sec:conclusion}

We have updated the modelling of the rapid evolution of Ly$\alpha$ emitters 
by CPBH15 \citep{2015MNRAS.452..261C} using the high-dynamic range Sherwood 
simulations as a basis for our analytical model for the growth of ionized regions.
We have in particular assessed the effect of host halo mass on LAE visibility 
just before the percolation of HII regions occurs at $z\sim 6$. Our main 
results can be summarised as follows:
\begin{itemize}

\item{Our simulations naturally reproduce the observed strong difference in
      the evolution of the visibility of bright and faint LAEs at $z\ga 6$ if 
      we assume that bright LAEs are placed in the most massive haloes in the 
      simulations with similar space densities as observed for bright LAEs.}
          
\item{The less rapid evolution of the visibility of bright LAEs in our 
      simulations at $z>6$ is only partially due to their strong spatial 
      correlation with the first regions to be reionized, an explanation that 
      has been invoked by other authors. In our simulations we find an
      additional contribution: the different gas peculiar/infall velocities 
      and peak temperatures in the environment of massive haloes contribute 
      to the differential evolution of bright and faint LAEs. 
      The relative contribution of the evolution of peculiar/infall 
      velocities and the spatial correlations with ionized regions on the 
      visibility of LAEs thereby depends strongly on the assumed reionization 
      history.} 
                   
\item{It is the faint emitters that more closely trace the evolution of the 
      volume-filling fraction of ionized regions, since the gas in their local 
      environments is not rapidly evolving (as it is for the bright emitters). 
      We thus recommend that studies of the reionization history continue 
      to focus on the fainter LAEs}. 

\item{In our simulations the infalling gas in the outskirts of the halo 
      (just outside the virial radius) has a strong effect on the visibility
      of the LAE it is hosting. This is in agreement with the suggestion by 
      \citet{2017ApJ...839...44S} that before percolation the infalling gas in 
      the outskirts of LAE host haloes in already ionized regions is  still 
      sufficiently neutral to cause a rapid evolution of LAE visibility at 
      $6<z<7$. In our simulations the photoionization rate in ionized regions 
      is higher than was modelled in that work, but the self-shielding is
      still sufficient to strongly attenuate the Ly$\alpha$ emission from the
      galaxy. In particular we find that this effect is stronger in the 
      more massive haloes. This means that for observations of UV bright 
      galaxies living in such massive hosts, deriving constraints on the
      volume-filling neutral fraction of the IGM involves more complicated
      modelling of such self-shielding than for UV faint LAEs living in
      less massive haloes. This reinforces our recommendation that future
      observational studies focus on UV faint LAEs for constraining 
      reionization. Alternatively, selecting LAEs based on intrinsic velocity
      offset could sample those galaxies whose emission is least attenuated by 
      the self-shielded gas of the CGM.}

\item{Overall our updated modelling with the higher dynamic range Sherwood 
      simulation gives similar results to CPBH15, albeit with some notable 
      differences:
      \begin{enumerate}

\item{We confirm that the `Late' and 'Very Late' reionization histories favoured
      in CPBH15, which also match Ly$\alpha$ forest data, are a good match to 
      the  observed 
      rapid evolution of faint  Ly$\alpha$ emitters. Note, however, that unlike 
      CPBH15  we can obtain this agreement without 
      invoking an  evolution of the redshift of the intrinsic Ly$\alpha$ 
      emission relative to systemic.  This is possibly due to the more 
      consistent treatment of peculiar velocities in our simulations made 
      possible by dropping the hybrid approach of CPBH15 (who combined 
      a rather small box-size hydrodynamical simulation with a large box-size 
      dark matter simulation). We further confirm that the evolution 
      of the ionizing emissivity  in the popular HM12 UV background model 
      corresponds to a decrease of the volume factor of ionized regions at 
      $z>6$ that is too slow to explain the rapid disappearance of faint LAEs. }

\item{As in CPBH15, in our updated simulations the rapid decrease of the 
      visibility of faint Ly$\alpha$ emitters is mainly due to the rapid 
      evolution of the volume-filling fraction of ionized regions in our models.  
      In our fiducial updated model we have used the  self-shielding  
      prescription  suggested by \citet{2017arXiv170706993C} who have explicitly 
      modelled the self shielding in ionized regions before the full 
      percolation of ionized regions with full radiative transfer simulations.  
      Note in particular that with this prescription the effect of 
      self-shielding is significantly 
      weaker than with the widely used \citet{2013MNRAS.430.2427R} model. If
      self-shielding is indeed as weak as suggested by the 
      \citet{2017arXiv170706993C} simulations, then reproducing the rapid 
      evolution of faint Ly$\alpha$ emitters at $z > 6$ may require a reionization
      history where reionization occurs as late as in our ``Very Late'' model.}

     \end{enumerate}
     }
\end{itemize}

The rapid disappearance of faint Ly$\alpha$ emitters arguably provides the 
strongest  constraints to date on the reionization history of hydrogen at 
$z>6$, and our simulations confirm that their rapid disappearance is strong 
evidence for a rather late reionization.

\section*{Acknowledgements}
We would like to thank Renske Smit and George Efstathiou for 
constructive comments, and Kazuaki Ota for useful 
discussion. We also thank an anonymous referee for helpful comments.
LHW is supported by the Science and Technology Facilities Council
(STFC). Support by ERC Advanced Grant 320596 `The Emergence of Structure 
During the Epoch of Reionization' is gratefully acknowledged.  We acknowledge
PRACE for  awarding  us  access  to  the  Curie  supercomputer,  based in
France  at  the  Tr\'{e}s  Grand  Centre  de  Calcul  (TGCC). This work used 
the DiRAC Data Centric system at Durham University,  operated  by  the 
Institute  for  Computational Cosmology  on  behalf  of  the  STFC  DiRAC  HPC
Facility (www.dirac.ac.uk). This equipment was funded by BIS National
E-infrastructure capital grant ST/K00042X/1, STFC capital  grants  ST/H008519/1
and  ST/K00087X/1,  STFC DiRAC Operations grant ST/K003267/1
and Durham University. DiRAC is part of the National E-Infrastructure.
This work made use of the \texttt{SciPy} \citep{SCIPY} ecosystem of libraries
for Python including: \texttt{NumPy} \citep{NUMPY}, \texttt{Matplotlib}
\citep{MATPLOTLIB} and \texttt{Cython} \citep{CYTHON}.


\bibliographystyle{mnras}
\bibliography{references}


\appendix

\section{Analytic modelling of halo infall profiles}
\label{sec:A}

\begin{figure*}
	\centering
    \begin{minipage}[c]{0.45\textwidth}
        \centering
        \includegraphics[width=\textwidth]{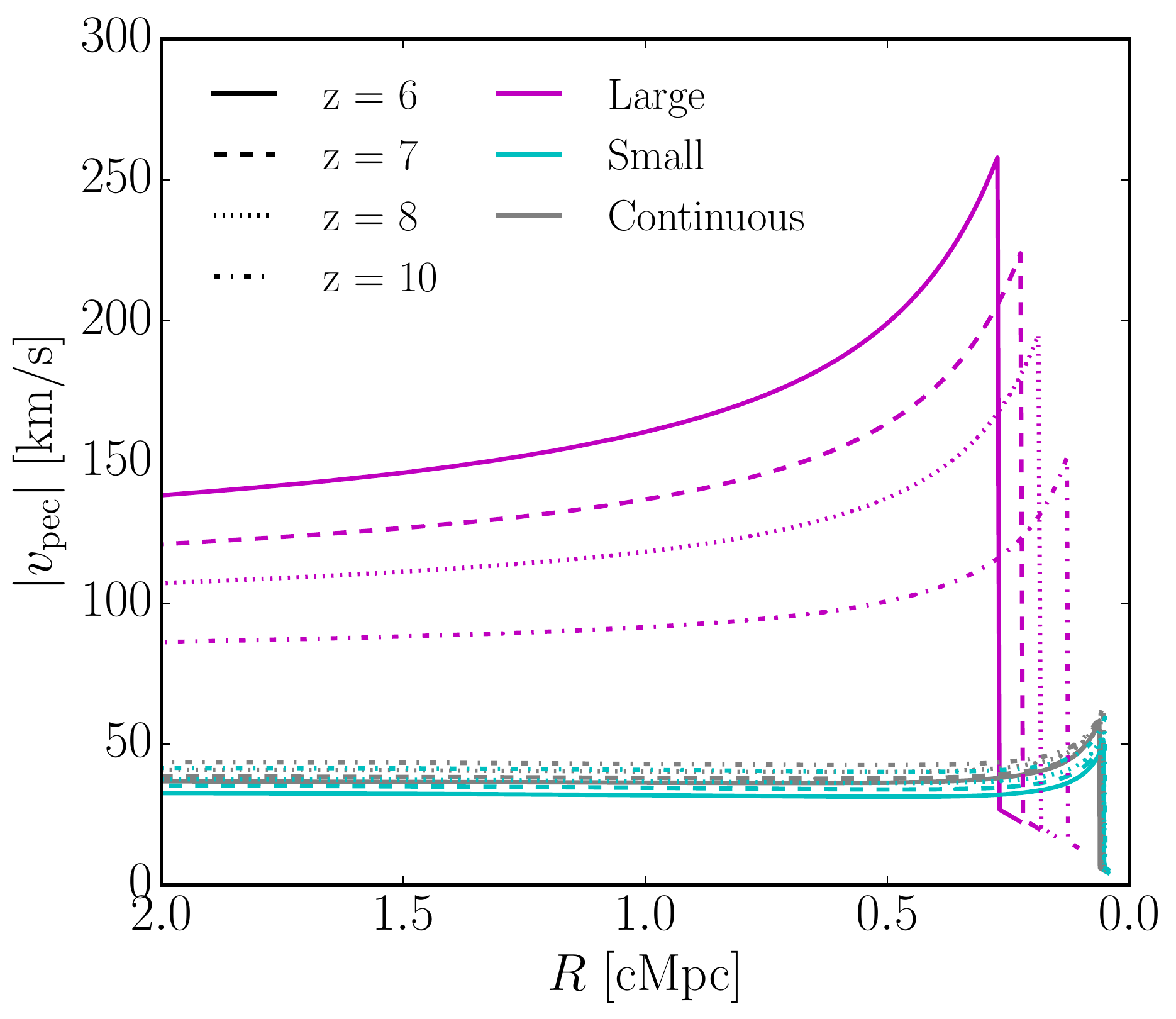}
    \end{minipage}
	\centering
    \begin{minipage}[c]{0.45\textwidth}
        \centering
        \includegraphics[width=\textwidth]{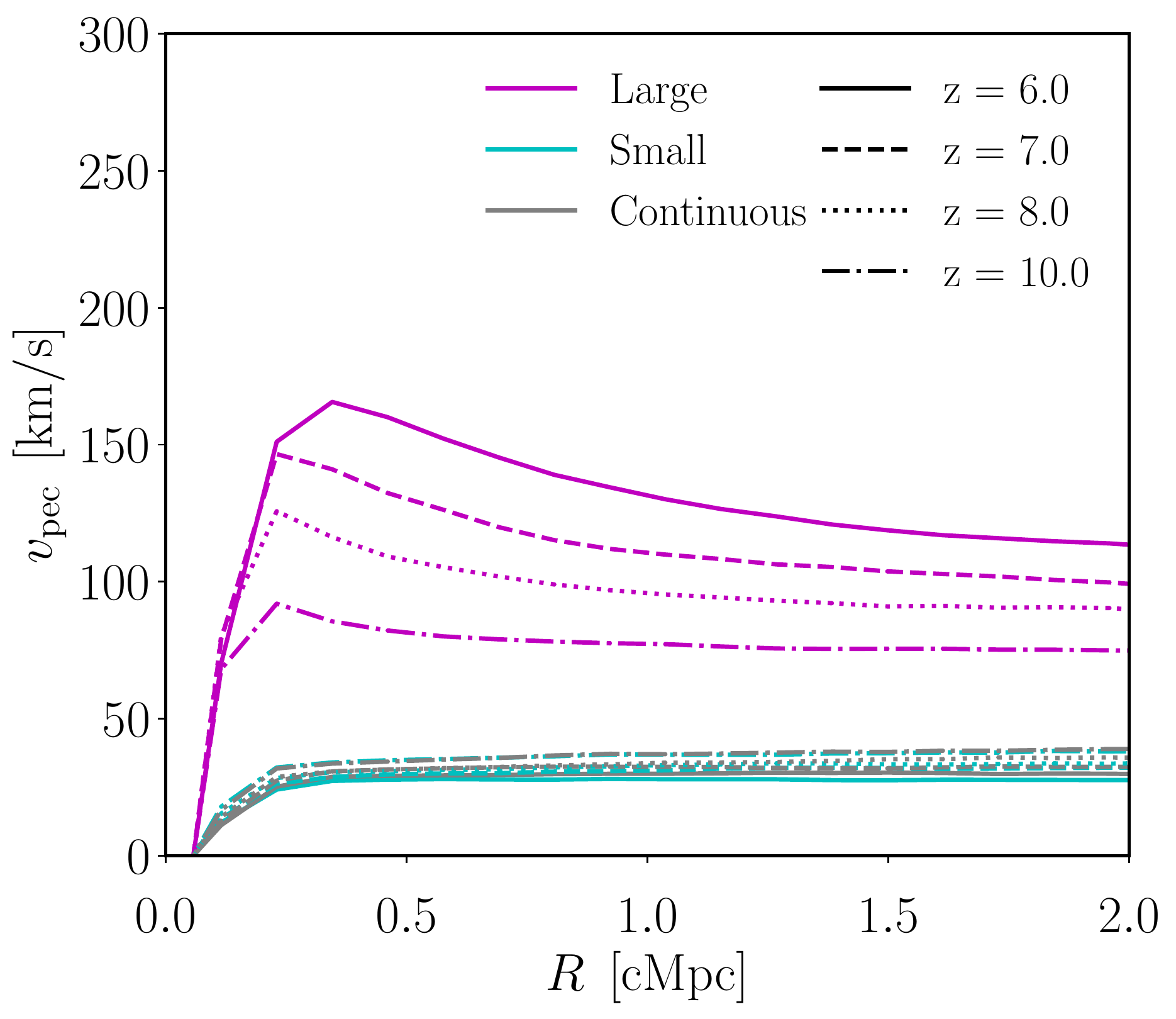}
    \end{minipage}
    \caption{Absolute values for the sightline 
        peculiar velocity of gas surrounding the  halo. 
    \emph{Left panel}: analytic predictions for the infalling velocity. 
    \emph{Right panel}: median velocity profiles from the simulation (where 
    the median is taken from the appropriately orientated sample of infall 
    velocities, and then the absolute value is taken). Note that the abscissa 
    has been mirrored about the vertical line between the panels, to aid
    comparison.}
    \label{fig:tirth_analytic}
\end{figure*}

\newcommand{\del}{\ensuremath{\partial}}
\newcommand{\Msun}{\ensuremath{{\rm M}_{\odot}}}
\newcommand{\avg}[1]{\ensuremath{\left\langle \,#1\, \right\rangle}}
\newcommand{\etal}{et al.}
\newcommand{\de}{\ensuremath{{\rm d}}}
\newcommand{\eqn}[1]{equation~\eqref{#1}}
\newcommand{\eqns}[1]{equations~\eqref{#1}}
\newcommand{\fig}[1]{Figure~\ref{#1}}
\newcommand{\figs}[1]{Figures~\ref{#1}}
\newcommand{\ph}[1]{\phantom{#1}}
\newcommand{\be}{\begin{equation}}
\newcommand{\ee}{\end{equation}}
\newcommand{\bear}{\begin{eqnarray}}
\newcommand{\ear}{\end{eqnarray}}
\newcommand{\nline}{\nonumber \\}
\newcommand{\f}{\frac}
\newcommand{\Cal}[1]{\ensuremath{\mathcal{#1}}}
\newcommand{\matrixsymbol}{\sf}

In this appendix, we discuss an analytical method for calculating the infall 
velocity profile around collapsed haloes. This is then applied to find the
velocity profiles we might expect to see around the average mass haloes of
our different mass models. We can then compare these profiles with the median profiles
in our simulation, as shown in Figure \ref{fig:sightlines}. Finally we construct
a simplistic model for the IGM gas surrounding an LAE using these analytic velocity
profiles; we then use this to calculate the difference in transmission due to
a differential velocity evolution. This differential transmission evolution is
similar to that found in the simulations as described in section \ref{sec:discussion}.

Our calculation closely follows that 
of \citet{2004MNRAS.347...59B} and \citet{2017ApJ...839...44S}, nevertheless we
summarize the main steps for completeness. The analytical calculation consists 
of two parts: 
\begin{enumerate}
    \item calculation of the 
linearly extrapolated initial density profile around the halo using the 
excursion set formalism;
    \item solving the non-linear problem for overdense 
spherical shells around the halo using the standard spherical collapse 
formalism.
\end{enumerate}

\subsection{Linearly extrapolated density profile}
Let us consider a halo of mass $M$ formed at some redshift $z$. In the language
of excursion sets, this problem can be mapped into a random walk problem in the
$s - \delta$ plane, where $s$ is the variance of the linearly extrapolated 
density contrast smoothed over some Lagrangian scale $r$ and $\delta$ is the 
linearly extrapolated smoothed density contrast at the same scale. Note that 
$s$, $r$ and the corresponding mass scale $m$ are related by the relations,
\be
s = \int_0^{\infty} \f{\de k}{k}~\f{k^3~P(k)}{2 \pi^2}~W^2(k r),~
m = \f{4 \pi}{3}~\bar{\rho}~r^3,
\ee
where $P(k)$ is the matter power spectrum linearly extrapolated to $z=0$, 
$\bar{\rho}$ is the present mean matter density of the universe and $W(k r)$ 
is the smoothing filter in Fourier space.

The formation of a halo corresponds to the first upcrossing of $\delta(s)$ of a threshold or 
a `barrier', $\delta_c(z)$, by random walks in the $s - \delta$ plane. In the spherical 
approximation, the barrier is independent of the scale $s$ and is given by 
$\delta_c(z) = 1.686 / D(z)$, $D(z)$ being the linear growth factor. The scale 
at which this upcrossing happens can be denoted by $s_M$ which typically falls near the
variance corresponding to the mass $M$. The linearly extrapolated density 
profile outside the halo can then be obtained from the distribution of 
$\delta(s)$ for $s \leq s_M$, with the condition that the random walks first 
upcross the barrier at $s_M$. It can be shown that the probability distribution
of the linearly extrapolated density profile can be written as,
\be
P(\delta, s | s_M) = \f{f(s_M | \delta, s)}{f(s_M)} ~ Q(\delta, s),
\ee
where $f(s_M)$ is the first upcrossing distribution, $f(s_M | \delta, s)$ is 
the \emph{conditional} first upcrossing distribution and $Q(\delta, s)$ is the 
probability that the walk has height $\delta$ at $s$ and remained below the 
barrier at all $s < s_M$.

In the case where the smoothing filter is chosen to be a tophat in $k$-space 
(i.e., the sharp-$k$ filter), the steps of the random walks become 
uncorrelated. In that case, we can write the above quantities as,
\bear
f(s_M) &=& \f{1}{\sqrt{2\pi}}~\f{\delta_c(z)}{s_M^{3/2}}~\exp\left(- 
\f{\delta_c^2(z)}{2 s_M} \right),
\nline
f(s_M | \delta, s) &=& \f{1}{\sqrt{2\pi}}~
\f{\delta_c(z) - \delta}{(s_M - s)^{3/2}}~\exp\left(- 
\f{[\delta_c(z) - \delta]^2}{2 (s_M - s)} \right),
\nline
Q(\delta, s) &=& \f{1}{\sqrt{2 \pi s}} 
\left[\exp\left(-\f{\delta^2}{2 s}\right) \right.
\nline
& & - \left. \exp\left(-\f{(2 \delta_c(z) - \delta)^2}{2 s}\right) \right].
\ear
The mean density profile is simply given by,
\be
\avg{\delta(s)}_M \equiv \int_{-\infty}^{\delta_c(z)} \de \delta ~ \delta
~ P(\delta, s | s_M),
\ee
where $s_M$ is set to the variance corresponding to mass $M$. This equation 
has the closed form solution \citep{2004MNRAS.347...59B},
\bear
\frac{\avg{\delta(s)}_M}{\delta_c(z)} &=& 1 - \left(1 - \alpha + \f{\alpha}{\beta}\right) 
{\rm erf}\left(\sqrt{\frac{\beta (1-\alpha)}{2 \alpha}}\right)
\nline
&-& \sqrt{\frac{2 \alpha (1 - \alpha)}{\pi \beta}} 
\exp\left(-\frac{\beta (1-\alpha)}{2 \alpha}\right),
\ear
where,
\be
\alpha \equiv \f{s}{s_M}, ~~\beta \equiv \f{\delta_c^2(z)}{s_M}.
\ee

\noindent Note that the calculation above assumes a sharp-$k$ filter for the 
random walks
but a real-space tophat filter for calculating the barrier height $\delta_c(z)$
\citep{1991ApJ...379..440B}. Removal of this inconsistency requires 
self-consistent usage of the real-space tophat filter for studying the random 
walks. However, this leads to steps that are correlated and are generally 
difficult to deal with. 
\citet{2004MNRAS.347...59B} proposed an ansatz based on the limit when the halo
corresponds to a rare peak $\beta \to \infty$. In that case, one can replace
the two parameters $\alpha$ and $\beta$ by the following
\be
\alpha \equiv \f{\xi(r_M, r)}{s_M},~~~
\beta \equiv \f{\delta_c^2(z)~\alpha~(1 - \alpha)}{s - \alpha \xi(r_M, r)},
\ee
where $r (r_M)$ is the Lagrangian length scale corresponding to $s (s_M)$, and
\be
\xi(r_M, r) = \int_0^{\infty} \f{\de k}{k} \f{k^3~P(k)}{2 \pi^2} 
~W(k r_M)~W(k r).
\ee
The quantity $W(x)$ is the $k$-space function corresponding to a spherical
tophat window in position space. We follow the above ansatz in our work as well.

\begin{figure}
	\centering
    \includegraphics[width=\columnwidth]{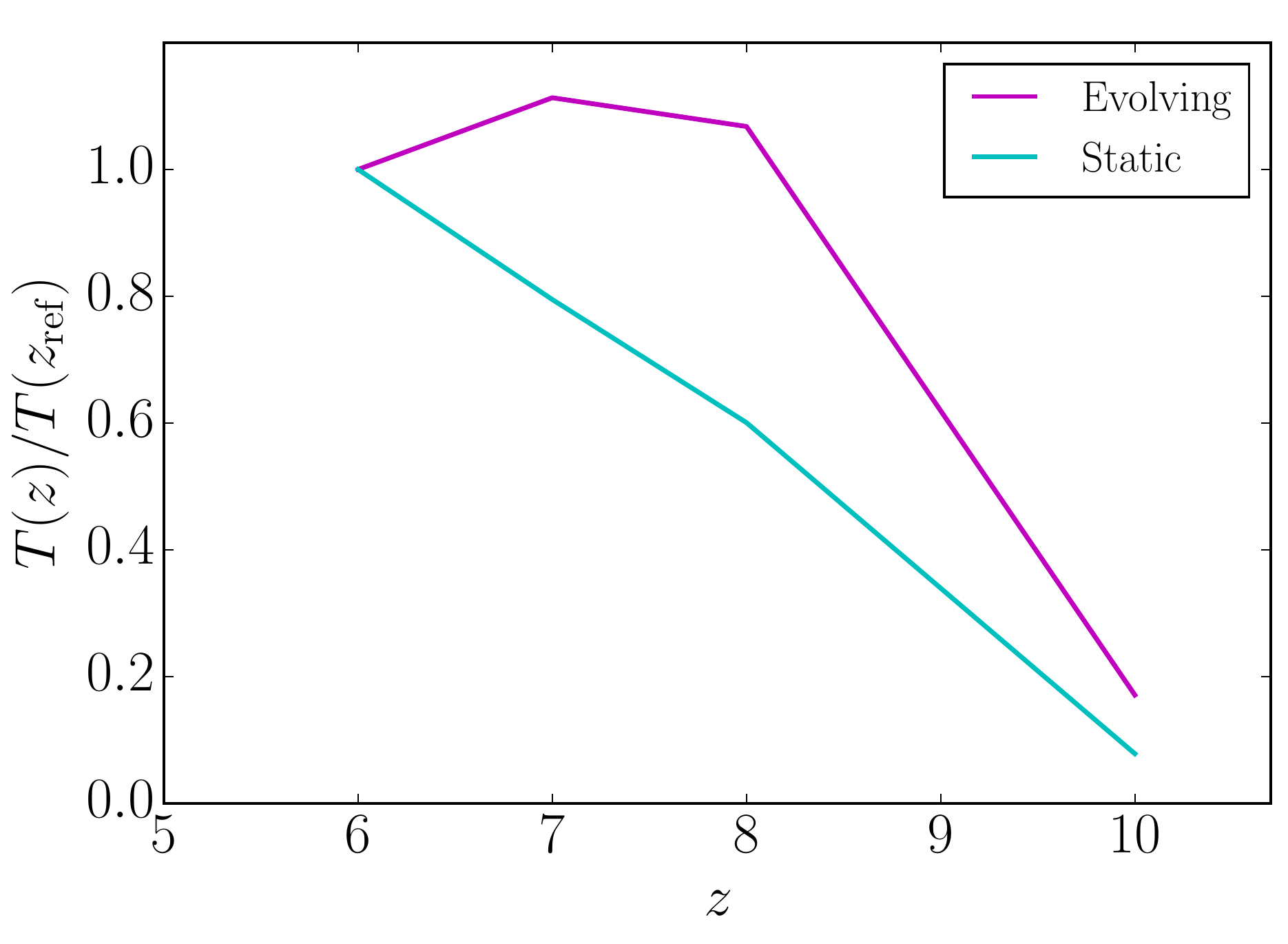}
    \caption{Mean transmission fraction evolution for the analytic models: 
    ``evolving'' and ``static'' with peculiar velocity structures based on 
    the large and small mass models as shown in Figure \ref{fig:tirth_analytic}.
    With only a difference in velocity structure, the mean transmission fraction
    evolution is changed dramatically, causing the ``evolving'' model LAEs to
    be more visible out to higher redshifts than the ``static'' LAEs.}
    \label{fig:simple_model_tfr}
\end{figure}

\subsection{Spherical collapse}

Now, consider the Lagrangian scale $r$ with mass $m$ and linearly extrapolated 
density contrast $\delta = \avg{\delta(s)}_M$. If we assume that the matter in a
region enclosed by $r$ evolves under spherical symmetry, we can apply the 
solutions of the spherical collapse directly. Since we are concerned with 
rather high redshifts $z \gtrsim 6$, we can work with the analytical solutions
obtained for the standard Einstein-deSitter universe.

The evolution of the spherical shell with linearly extrapolated density 
contrast $\delta$ and enclosed mass $m$ is given by the parametric solution 
\citep[see, for example,][section 5.1.1]{2010gfe..book.....M}
\bear
R = A~(1 - \cos \theta),&& t = B~(\theta - \sin \theta),
\nline
B = \f{6 t_i}{(20~\delta_i / 3)^{3/2}}, && A^3 = G m B^2,
\label{eq:spherical_collapse}
\ear
where $t_i$ is the initial time and $\delta_i = D(t_i)~\delta$. Note that the 
radius $R$ in the above solution is in proper units. The velocity of the shell
is then
\be
v = \f{A}{B}~\f{\sin \theta}{1 - \cos \theta}.
\ee
The peculiar velocity is obtained by subtracting the Hubble velocity $H(t)~R$ 
from the above, i.e.,
\be
v_{\rm pec} = v - H(t)~R.
\ee
Note that $v_{\rm pec} > 0$ for outflowing or expanding matter. Since we are
interested in the infall velocity around the haloes, we shall plot the 
magnitude of the velocity, $|v_\mathrm{pec}|$, to compare between the
analytic model and the simulations.

The above solution breaks down once tangential motions and shell crossings
become important. In that case, the collisionless dark matter virializes via
violent relaxation while the baryons undergo various non-linear processes which
lead to the formation of galaxies. Since such non-linear processes are difficult 
to account for, usually one assumes that the time of virialization (or 
collapse) is that corresponding to $\theta = 2 \pi$, and the final virial 
radius is given by the virial energy condition which turns out to be
\be
R_{\rm vir} = A.
\ee
According to the solution (\ref{eq:spherical_collapse}), the value of $R$ 
approaches $R_{\rm vir}$ when $\theta = 3 \pi / 2$. Clearly, the spherically
symmetric solution has crossed the regime of validity by then. We make the 
simplifying assumption that the radius remains constant for 
$\theta > 3 \pi / 2$
\be
R = \left\{\begin{array}{ll}
A~(1 - \cos \theta) & \displaystyle \mbox{when } \theta \leq \f{3 \pi}{2},\\
&\\
A & \mbox{otherwise}.
\end{array}\right.
\label{eq:radius}
\ee
Correspondingly, the radial velocity is assumed to be
\be
v = \left\{\begin{array}{ll}
\displaystyle \f{A}{B}~\f{\sin \theta}{1 - \cos \theta} & \displaystyle 
        \mbox{when } \theta \leq \f{3 \pi}{2},\\
&\\
0 & \mbox{otherwise}.
\end{array}\right.
\label{eq:velocity}
\ee
The above relation assumes that as the halo approaches virialization, 
the radial velocity becomes zero and all the kinetic energy has been 
converted into random motions.

The two \eqns{eq:radius} and (\ref{eq:velocity}) together give the velocity
profile which can be directly compared with simulation results. The calculation
is strictly meant for collisionless matter, however, we apply it to the gas 
profile under the assumption that the gas follows the dark matter outside the 
virial radius \citep[as in][]{2017ApJ...839...44S}.

\subsection{Comparison with simulation velocity profiles}
We apply this formalism to calculate the velocity profiles for masses 
corresponding to the average masses of our models. These masses are shown for
$z$ = 6, 7, 8 \& 10 in Table \ref{tab:models}. The resulting profiles for the
large, small and continuous average mass haloes can 
be seen in Figure \ref{fig:tirth_analytic}.

In Figure \ref{fig:tirth_analytic} we see the same behaviour observed in the
median profiles from the simulation. In particular, we see a similar redshift 
evolution of the profiles across the models: in the large model we see an 
increase in infall velocities with decreasing redshift; conversely in the small
and continuous models we see a much smaller evolution (and in the opposite 
direction -- peak velocity increasing with redshift). 
This effect plays a role in the different evolution of the 
visibility of the models, as discussed in section \ref{sec:discussion}.

\subsection{The effect of infall velocity evolution}
In order to test whether this infalling velocity evolution can lead to 
a differential visibility evolution we construct two simple models for the gas 
properties around the host halo of an LAE\@. We keep the neutral hydrogen density,
$n_\mathrm{HI}$, and gas temperature, $T_\mathrm{HI}$, the same in both models; we then
use the velocity profiles calculated with Eq.~(\ref{eq:velocity}) (seen in Figure
\ref{fig:tirth_analytic}) to construct an ``evolving'' velocity profile and a
``static'' profile. The ``evolving'' profile has a larger infalling velocity 
amplitude, as well as a strong evolution with redshift, and it therefore presents
a similar gas environment to that around the large mass model haloes. In comparison
for the ``static'' profile we fixed the velocity profile to be constant with redshift,
and with a lower infalling amplitude, as is the case for the environments around
the small mass model haloes. 

In order to create a similar macroscopic evolution
of the Ly$\alpha$ transmission as we see in the simulations we set the neutral
hydrogen fraction around the LAE to be either an equilibrium value 
$x_\mathrm{HI}=x_\mathrm{eq}$ (found using $\Gamma_\mathrm{HI}$) 
or $x_\mathrm{HI}=1$ (for regions not yet reionized), based on $Q_\mathrm{M}$.
We use $Q_\mathrm{M}$ and $\Gamma_\mathrm{HI}$ from the Late reionization history.
The temperature of the gas is fixed at $T_\mathrm{HI} = 10^4$ K, and the
total hydrogen density is chosen to be the mean cosmic hydrogen density. 
This crude modelling
of $n_\mathrm{HI}$ and $T_\mathrm{HI}$ is intended as a zeroth order description
the IGM gas, and importantly is the same for each case (``evolving'' or
``static''). Given these gas properties we calculate the Ly$\alpha$ transmission
as in section \ref{sec:calculations}.

The resulting mean transmission fraction (TFR) evolution is shown in 
Figure \ref{fig:simple_model_tfr}, for the ``evolving'' profile in magenta and the
``static'' model in cyan. We see that the ``evolving'' profile results in a slower
evolution of the transmission. 

We note that the presence of this infalling gas velocity evolution can 
therefore lead to a differential visibility in LAEs. The magnitude of this
difference in the visibility is dependent on the neutral gas density; in reality
the profiles close to the host halo will differ significantly from the crude model
discussed above.

This effect can be understood as follows: when the infalling velocities are
comparable to the instrinsic offset of the emission, then neutral gas close to
the emitter will strongly absorb Ly$\alpha$. As these velocities decrease
with increasing redshift, there will be a decrease in the absorption from this
self-shielded CGM gas. This decrease in absorption acts counter to the
increase in absorption in the neutral gas of the larger-scale IGM, which is
increasing with redshift (as less reionization has ocurred). Hence this
velocity structure can counter some of the transmission evolution, and 
will -- despite itself being due to a \emph{rapid} evolution of the velocity amplitude 
-- \emph{slow} the transmission evolution, in the more massive haloes where it can be
significant.

\section{Further model variations}
\label{sec:C}

\begin{figure*}
    \begin{minipage}{0.49\textwidth}
   	\includegraphics[width=\textwidth]{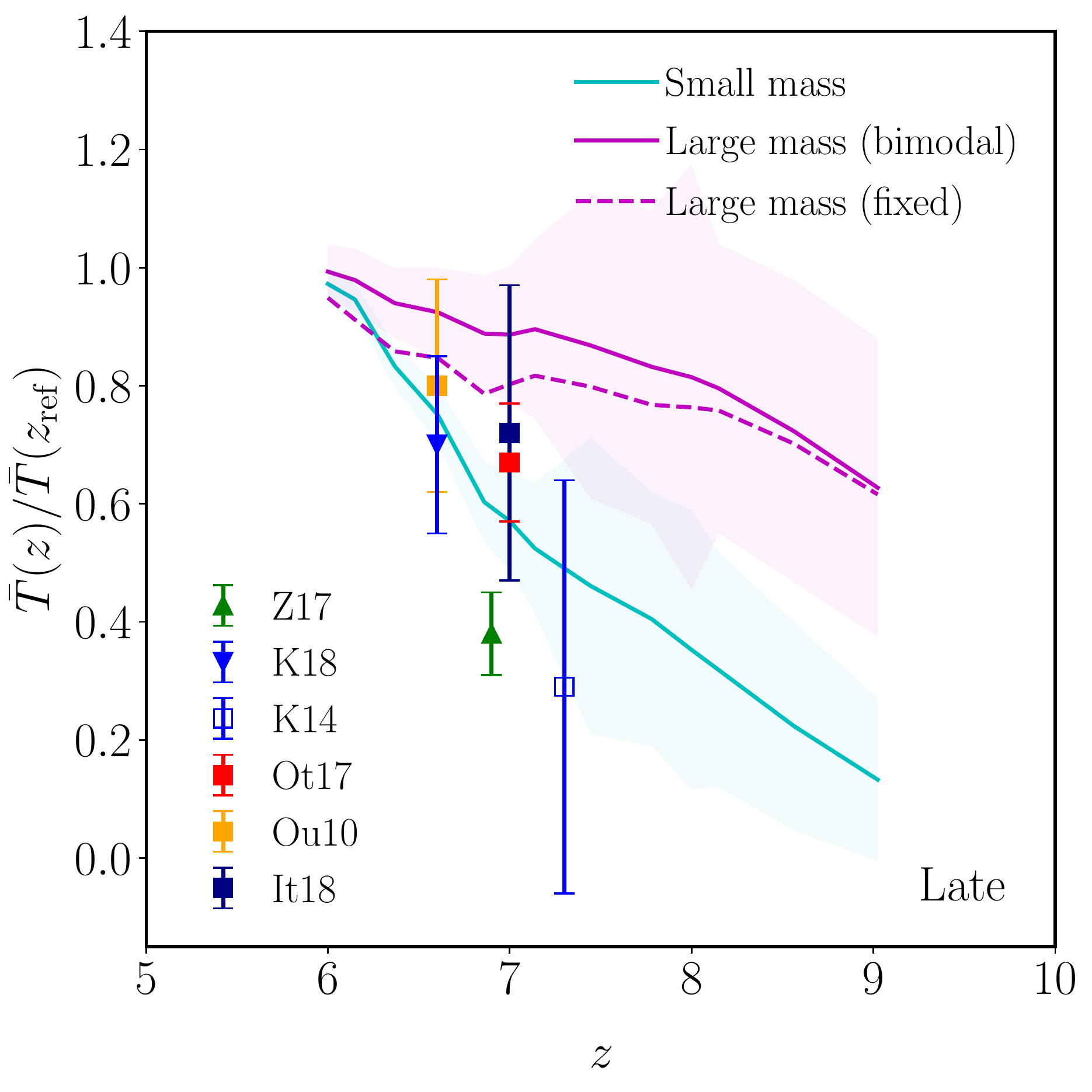}
    \end{minipage}
    \begin{minipage}{0.49\textwidth}
  	\includegraphics[width=\textwidth]{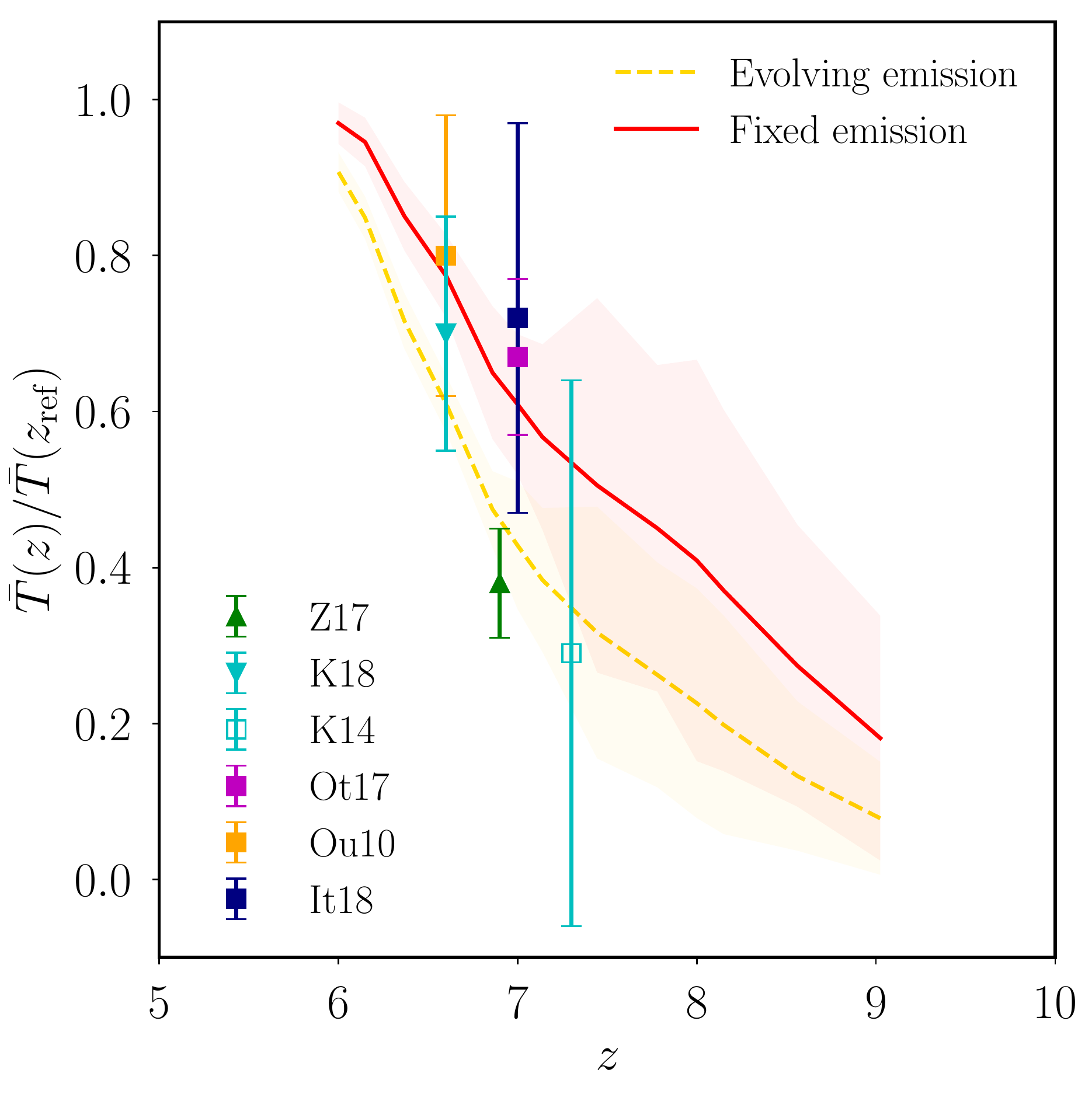}
    \end{minipage}
    \caption{
    \emph{Left panel}:
    The TFR evolution for the large and small mass models, using the
    mass dependent ``bimodal'' emission profile. For comparison the default
    emission profile (``fixed'') is shown as a dashed line.
    Overplotted are observed TFR values, as in Figure \ref{fig:ratio_evol}.
    \emph{Right panel}: 
    TFR evolution in the continuous model with the Late reionization
    history: comparing a redshift-dependent emission profile to the default
    fixed profile. The evolving velocity offset leads to an increased Ly$\alpha$ 
    attentuation.
}
    \label{fig:other_emission1}
\end{figure*}

\noindent We have also explored further variations to the 9 model combinations
presented in this work. In particular we vary the emission profile model
from our fiducial, by considering different intrinsic velocity offsets.

The default model in this work assumed a Gaussian emission profile, with 
width $\sigma_v$ = 88 km/s and line-centre offset from Ly$\alpha$ by
$\Delta v_\mathrm{int}$ = 100 km/s. This model tries to account for the
complex radiative transfer within the galaxy that leads to a reddened peak.
We applied this model to all haloes, regardless of mass.

We now test a second model using a bimodal distribution of profiles: for the 
small mass range we use the default
$\Delta v_\mathrm{int} = 100\:\mathrm{km \: s^{-1}}$, but for the
large mass range we use a larger offset of 
$\Delta v_\mathrm{int} = 300\:\mathrm{km \: s^{-1}}$. 
This bimodal model is motivated by some recent observational results, for 
example \citet{2015ApJ...807..180W} and \citet{2017MNRAS.464..469S}, which 
have found that the most luminous LAEs at $z\sim6$ can have 
$\Delta v_\mathrm{int} =$ 300 -- 500 $\mathrm{km \: s^{-1}}$. 
We also test a third model, which varies the velocity offset as a function of 
redshift,
\begin{equation}
    \Delta v_\mathrm{int} = 100\left(\frac{1+z}{7}\right)^{-3} 
    \:\: \mathrm{km \: s^{-1}}
    \label{eq:voff}
\end{equation}
This model was employed in CPBH15 to explore further enhancements to the
IGM absorption, and was found to aid the agreement with the data.

In  the left panel of Figure \ref{fig:other_emission1} we show the TFR evolution for
the large mass and small mass models using the mass-dependent profile. The evolution
of the large model is changed slightly, but not dramatically. Although this
bimodal profile can lead to much higher transmission fractions in the large mass
model, it does so for all redshifts, and hence the TFRs (which are ratios 
across redshifts) are broadly unaffected. We note however that the scatter in 
the TFR is greatly reduced compared to the fixed emission profile TFRs.

In the right panel of Figure \ref{fig:other_emission1} we show the
TFR evolution for the continuous model with the redshift-dependent emission
profile. Here we see an increase in attenuation which increases the agreement 
between the data and the `Late' reionization history. If indeed
the velocity offset
of LAEs evolves in such a manner, this would reinforce the conclusion that
reionization progressed in a `Late' history.

We therefore find that our main conclusions regarding the differential
evolution of different mass LAE host halos and the best fitting reionization
history are robust to these changes in emission profile.

\bsp
\label{lastpage}
\end{document}